\documentclass[%
 aip,
 jcp,
 amsmath,amssymb,
 reprint,%
]{revtex4-2}
\usepackage{xr} 
\usepackage{hyperref}
\hypersetup{
    colorlinks=true,
    linkcolor=blue,
    citecolor=blue,
    filecolor=blue,      
    urlcolor=blue,
    }
\usepackage{graphicx}
\usepackage{dcolumn}
\usepackage{bm}
\usepackage[dvipsnames]{xcolor}

\usepackage[utf8]{inputenc}
\usepackage[T1]{fontenc}
\usepackage{mathptmx}
\usepackage{etoolbox}

\usepackage{float}

\makeatletter
\newcommand*{\addFileDependency}[1]{
\typeout{(#1)}
\@addtofilelist{#1}
\IfFileExists{#1}{}{\typeout{No file #1.}}
}

\makeatother

\newif\ifhighlight
\highlighttrue   
\highlightfalse 

\ifhighlight
  \newcommand{\blue}[1]{{\color{blue}#1}}
\else
  \newcommand{\blue}[1]{{#1}}
\fi

\makeatletter
\def\@email#1#2{%
 \endgroup
 \patchcmd{\titleblock@produce}
  {\frontmatter@RRAPformat}
  {\frontmatter@RRAPformat{\produce@RRAP{*#1\href{mailto:#2}{#2}}}\frontmatter@RRAPformat}
  {}{}
}%
\makeatother
\begin{document}

\preprint{AIP/123-QED}

\title{Imaging transient molecular configurations in UV-excited diiodomethane}
\author{Anbu Selvam Venkatachalam}
\email{anbu@ksu.edu}
\affiliation{James R. Macdonald Laboratory, Physics Department, Kansas State University, Manhattan, KS 66506, USA}
\author{Huynh Van Sa Lam}
\affiliation{James R. Macdonald Laboratory, Physics Department, Kansas State University, Manhattan, KS 66506, USA}
\author{Surjendu Bhattacharyya}
\thanks{Current address: SLAC National Accelerator Laboratory, Menlo Park, CA 94025, USA.}
\affiliation{James R. Macdonald Laboratory, Physics Department, Kansas State University, Manhattan, KS 66506, USA}
\author{Balram Kaderiya}
\affiliation{James R. Macdonald Laboratory, Physics Department, Kansas State University, Manhattan, KS 66506, USA}
\author{Enliang Wang}
\thanks{Current address: Department of Modern Physics, University of Science and Technology of China, Anhui, China}
\affiliation{James R. Macdonald Laboratory, Physics Department, Kansas State University, Manhattan, KS 66506, USA}
\author{Yijue Ding}
\thanks{Current address: Department of Chemistry, Southern University of Science and Technology, Shenzhen, Guangdong 518000, China}
\affiliation{James R. Macdonald Laboratory, Physics Department, Kansas State University, Manhattan, KS 66506, USA}
\author{Loren Greenman}
\affiliation{James R. Macdonald Laboratory, Physics Department, Kansas State University, Manhattan, KS 66506, USA}
\author{Artem Rudenko}
\affiliation{James R. Macdonald Laboratory, Physics Department, Kansas State University, Manhattan, KS 66506, USA}
\author{Daniel Rolles}
\email{rolles@ksu.edu}
\affiliation{James R. Macdonald Laboratory, Physics Department, Kansas State University, Manhattan, KS 66506, USA}

\date{\today}

\begin{abstract}
Femtosecond structural dynamics of diiodomethane ($\mathrm{CH_{2}I_{2}}$) triggered by ultraviolet (UV) photoabsorption at 290 nm and 330 nm are studied using time-resolved coincident Coulomb explosion imaging driven by a near-infrared probe pulse. We map the dominant single-photon process, the cleavage of the carbon-iodine bond producing rotationally excited $\mathrm{CH_{2}I}$ radical, identify the contributions of the three-body ($\mathrm{CH_{2} + I + I}$) dissociation and molecular iodine formation channels, which are primarily driven by the absorption of more than one UV photon, and demonstrate the existence of a weak reaction pathway involving the formation of short-lived transient species resembling iso-CH$_{2}$I$_{2}$-like geometries with a slightly shorter I--I separation compared to the ground-state $\mathrm{CH_{2}I_{2}}$. These transient molecular configurations, which can be separated from the other channels by applying a set of conditions on the correlated momenta of three ionic fragments, are formed within approximately 100 fs after the initial photoexcitation and decay within the next 100 fs.
\end{abstract}

\maketitle

\section{\label{sec:level1}Introduction}

The interaction between light and matter is a common and essential aspect of many fundamental processes in nature\cite{Kamat2015PhotonsIssue, Glusac2016WhatChemistry}, including photosynthesis\cite{Govindjee1974ThePhotosynthesis, Renger2010ThePhotosynthesis, Zubik2012TheState}, vision\cite{Schoenlein1991TheRhodopsin, Peteanu1993TheStudies., Wang1994VibrationallyVision, Polli2010ConicalVision}, vitamin D synthesis\cite{Holick1980PhotosynthesisConsequences, Maclaughlin1982SpectralSkin, HOLICK1988SkinD3}, DNA repair\cite{Sinha2002UV-inducedReview, Essen2006Light-drivenPhotolyases, Gustavsson2010DNA/RNA:Irradiation}, and various atmospheric reactions\cite{Keller-Rudek2013TheInterest}. Among the latter, UV-induced photodissociation reactions of halogenated alkanes are significant sources of reactive halogens, which have a considerable impact on environmental and atmospheric chemistry. Iodine, one of the halogens, plays diverse roles in chemistry, serving as a fundamental element in human health and bio-chemistry, a useful catalyst in organic synthesis \cite{Schomburg2008OnHealth}, and a major contributor to the destruction of ozone molecules \cite{Koenig2021OzoneTroposphere}. Due to its strong absorption of sunlight in a broad range of UV wavelengths, diiodomethane ($\mathrm{CH_{2}I_{2}}$), a polyhalogenated alkane, is a major source of highly reactive iodine molecules influencing tropospheric chemistry and the marine boundary layer\cite{Carpenter2003IodineLayer}. 

Numerous studies have been published on the UV-induced photochemistry of $\mathrm{CH_{2}I_{2}}$, which have reported the primary cleavage of one of the C--I bonds and the formation of $\mathrm{CH_{2}I}$ and $\mathrm{I (^{2}P_{3/2})}$ or $\mathrm{I^{*} (^{2}P_{1/2})}$ photoproducts \cite{Kawasaki1975PhotodissociationIodoform, Kroger1976PolyhalideIodide, Baughcum1980PhotofragmentationCH2I}.
Using ions generated by a (2+1) resonance enhanced multiphoton ionization process, Xu \textit{et al.}\cite{Xu2002PhotodissociationImaging} measured the translational kinetic energy distributions of both $\mathrm{I (^{2}P_{3/2})}$ and $\mathrm{I^{*} (^{2}P_{1/2})}$ fragments in the wavelength range of 277--305 nm and concluded that the $\mathrm{CH_{2}I}$ co-fragment is produced with significant internal excitation, with approximately 80\% of the total available energy being partitioned into the internal energy of the $\mathrm{CH_{2}I}$ fragment.

Reid \textit{et al.}\cite{Kalume2010IsomerizationHalons, Reid2014WhenIso-halocarbons} later drew attention to the importance of isomerization of halogenated alkanes into iso-haloalkanes upon UV absorption as a major pathway leading to the production of molecular halogens. The photoisomerization of $\mathrm{CH_{2}I_{2}}$ to iso-$\mathrm{CH_{2}I_{2}}$ has been reported to be an extremely efficient process upon UV absorption in the solution phase and in cages\cite{Tarnovsky2004PhotochemistryOlefins,Panman2020ObservingScattering, Kim2019FateCH2I2}. 
Since it has been suggested that the dominant isomerization mechanisms in these studies are driven by the interaction with solvent molecules or cage, it is helpful to examine the process in isolated gas-phase molecules, free from such interactions. Borin \textit{et al.}\cite{Borin2016DirectStudy} reported observing photoisomerization of $\mathrm{CH_{2}I_{2}}$ in the gas phase upon 330 nm UV excitation by probing it with femtosecond transient absorption. 
The delayed rise in the transient absorption of $\mathrm{CH_{2}I_{2}}$ at 380 and 612~nm, approximately 35--90 fs after UV excitation, was attributed to the creation of a short-lived iso-$\mathrm{CH_{2}I_{2}}$. 
A recent ultrafast electron diffraction (UED) study, where Liu \textit{et al.}\cite{Liu2020SpectroscopicDiffraction} directly mapped the C--I cleavage upon 266 nm excitation, quantified the resulting increase of the C--I and I--I separations, and observed signatures of the rotation of the $\mathrm{CH_{2}I}$ radical after the C--I bond cleavage, but did not mention any observation of iso-$\mathrm{CH_{2}I_{2}}$ formation.

Here, we employ Coulomb Explosion Imaging (CEI), a powerful method for determining the geometric structure of gas-phase molecules \cite{Vager1989CoulombMolecules, Kella1993AInstitute, Cornaggia2009UltrafastMolecules, Corrales2018CoulombIntersection, Ablikim2016IdentificationImaging, Pathak2020DifferentiatingImaging, Boll2022X-rayMolecules, Bhattacharyya2022Strong-Field-InducedTribromomethane, Li2022CoulombPulses, Lam2023CoulombReactions, Wang2023, Lam2024DifferentiatingImaging, Jahnke2025, Venkatachalam2025DCEML, Xiang2025CH2I2Imaging}, to investigate the structural dynamics of $\mathrm{CH_{2}I_{2}}$ upon UV photoabsorption, specifically at 290~nm and 330~nm.
Excitation at both 290 nm (4.28 eV) and 330 nm (3.76 eV) can access several excited states that are adiabatically leading to CH$_2$I+I($^2$P$_{3/2}$) dissociation. 
Excitation at 290 nm can also access states leading to CH$_2$I+I$^*$($^2$P$_{1/2}$) dissociation\cite{Toulson2016Near-UVCH2I2}.
CEI has been shown to be a useful tool in identifying molecular isomers \cite{Ablikim2016IdentificationImaging} and conformers \cite{Pathak2020DifferentiatingImaging} of polyhalogenated alkanes and other organic molecules \cite{Lam2024DifferentiatingImaging}, and even imaging the complete structure of halogenated alkanes \cite{Bhattacharyya2022Strong-Field-InducedTribromomethane, Li2022CoulombPulses} and ring molecules \cite{Boll2022X-rayMolecules, Lam2023CoulombReactions, Lam2024DifferentiatingImaging}. When used as a time-resolved method in a pump-probe scheme, laser-induced CEI can measure wave-packets dynamics\cite{Stapelfeldt1995WaveExplosion, Lam2020PRA, Lam2025PRAL}, visualize vibrational motions\cite{Stapelfeldt1998Time-resolvedPackets, Ergler2006UltrafastImaging, Lam2025PRAL}, and image photo-dissociation\cite{Bocharova2011Time-resolvedPulses, Amini2018PhotodissociationIonization, Ziaee2023Single-and, Bhattacharyya2024JPCL} 
with high temporal 
resolution\cite{Legare2005ImagingDynamics}. In particular, CEI provides multidimensional data crucial for identifying any transient structural changes in the molecule upon UV excitation, often with the help of simple classical modeling and computations.
\begin{figure}
\includegraphics[width=\columnwidth]{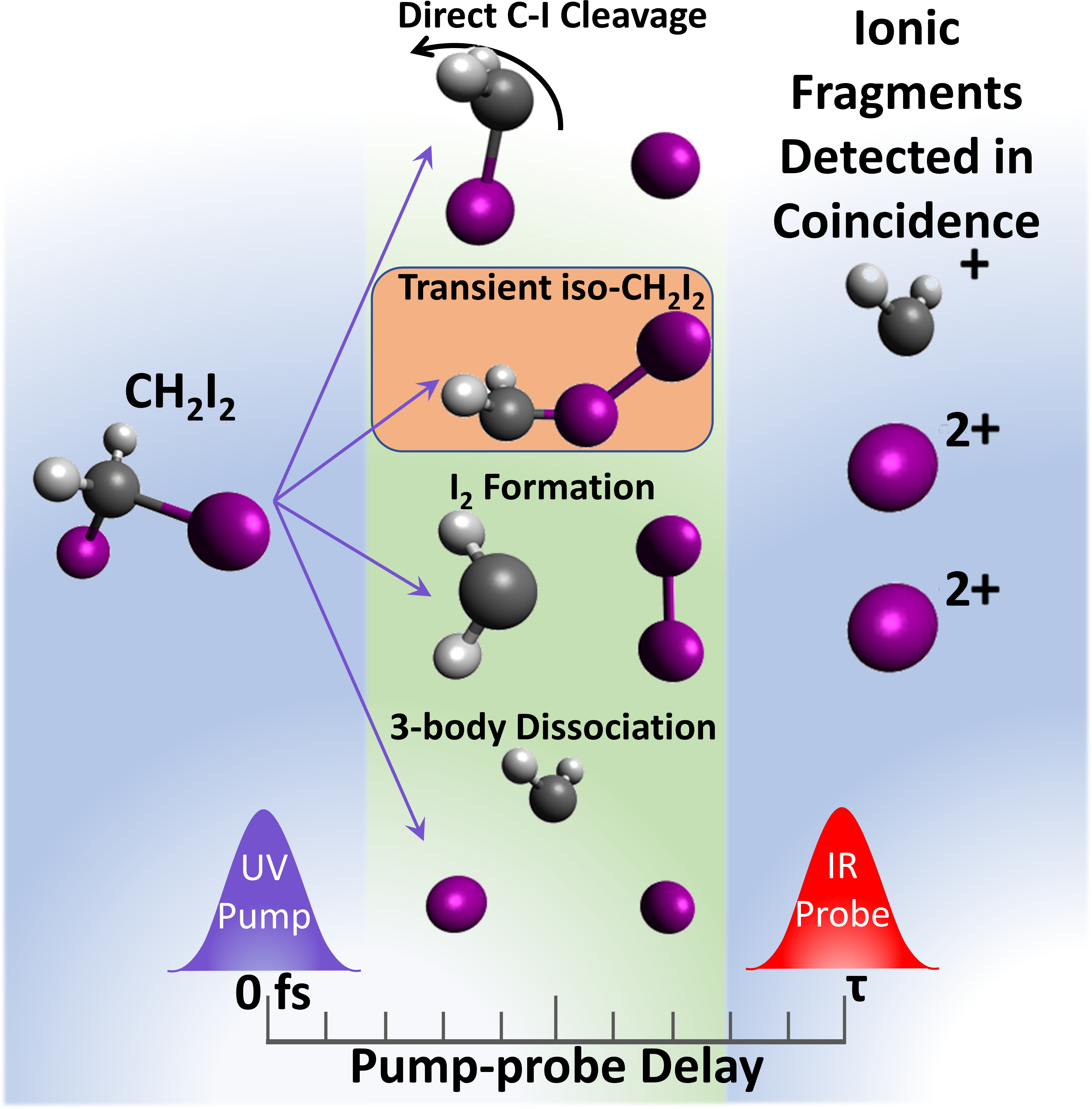}
\caption{\label{fig:schematic} Sketch depicting the pump-probe process and the different reaction pathways after photoexcitation of $\mathrm{CH_{2}I_{2}}$ (left) by the ultraviolet (UV) pump pulse. The products and possible intermediates (middle) are ionized to five-fold final charge state by the intense near-infrared (NIR) probe pulse, and the ionic fragments are detected in coincidence.}
\end{figure}

In this work, we investigate the dynamics of diiodomethane upon UV excitation using ionization and Coulomb explosion by an intense strong-field near-infrared (NIR) probe pulse. The primary objective of this time-resolved study is to explore possible signatures of the intramolecular photoisomerization process in $\mathrm{CH_{2}I_{2}}$, guided by the findings reported by Borin \textit{et al.} \cite{Borin2016DirectStudy}. We analyze the delay-dependent angular correlations between the momenta of the ions, the total kinetic energy release (KER), and the kinetic energies (KE) of individual ionic fragments to differentiate the possible photoisomerization channel from the direct two-body breakup channel that leads to $\mathrm{CH_{2}I + I}$ (or $\mathrm{I^{*}}$) products. We also identify the three-body ($\mathrm{CH_{2} + I + I}$) dissociation and the I$_2$ formation channels, which primarily occur after the absorption of multiple UV photons. A schematic of the pump-probe experiment and the different reaction pathways after UV photoexcitation is shown in Fig.~\ref{fig:schematic}.

\section{Methods}
\subsection{Experimental}
    
\begin{figure}
\includegraphics[width=\columnwidth]{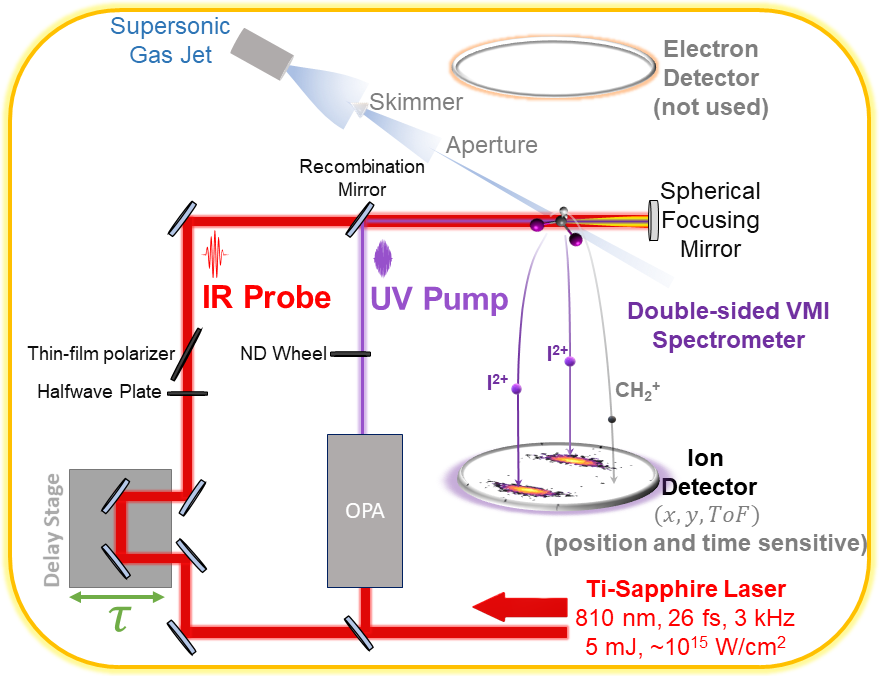}
\caption{\label{fig:ExpSetup} Schematic of the pump-probe setup and the coincident ion momentum imaging system. The diagram depicts the propagation of the ultraviolet (UV) pump and near-infrared (NIR) probe pulses into the experimental chamber. Both pulses are collinearly directed and focused onto a cold supersonic molecular jet containing diiodomethane. The setup is integrated with a double-sided velocity map imaging (VMI) spectrometer, operated in multi-ion coincidence mode. Ions are detected in coincidence using a time- and position-sensitive detector comprising a microchannel plate (MCP) and delay line anode assembly. It should be noted that in this particular experiment, the electron detector is not utilized.}
\end{figure}

Figure~\ref{fig:ExpSetup} illustrates the experimental setup for the UV-pump and NIR-probe CEI experiment. The femtosecond laser system comprises a 5-mJ, 3-kHz Coherent Legend Elite DUO Ti:Sapphire laser coupled to a Light Conversion TOPAS Prime optical parametric amplifier (OPA). The output of the Ti:Sapphire laser is split evenly into two independently compressed beam paths: one driving the OPA, and the other providing near-Fourier-transform-limited NIR pulses with a full width at half maximum intensity (FWHM) duration of 26~fs and a bandwidth of 60~nm at a central wavelength of 810~nm. In this experiment, the OPA was set to produce pulses at a central wavelength of 290~nm and 330~nm, with a 5-nm bandwidth, pulse duration of 70~fs (FWHM), and pulse energy up to 15~${\mu}$J. The UV and NIR pulses were recombined using a 45-degree recombination mirror that transmitted the NIR pulses and reflected the UV pulses, and the combined beam was directed collinearly into the vacuum chamber through a 1-mm calcium fluoride $\mathrm{(CaF_{2})}$ window. The delay between NIR and UV pulses was adjusted via a computer-controlled optical delay stage in the NIR arm. Inside the vacuum chamber, the laser pulses are focused using a normal-incidence spherical mirror with 75-mm focal length and UV-enhanced aluminum coating. 

Diiodomethane, which is liquid at room temperature, is introduced into the vacuum chamber as a supersonic molecular beam expanded through a 30-${\mu}$m nozzle using helium at 3 psig as carrier gas, which is collimated by a skimmer with a 500-${\mu}$m diameter opening. A second differential pumping stage in the molecular beam, separated from the interaction chamber by a 700-${\mu}$m aperture, and a beam dump comprising of two differential pumping stages ensure that the base vacuum in the interaction chamber stays around $\mathrm{10^{-10}}$ mbar when the molecular beam is operating.

Ions produced by the interaction between the focused laser pulses and the molecular beam are collected using a double-sided velocity map imaging (VMI) spectrometer. This apparatus is similar to the one employed in the CAMP end-station for free-electron laser experiments \cite{Rolles2014FemtosecondMolecules, Erk2018CAMPFLASH:Laser}, while using position-sensitive delay-line detectors (Roentdek DLD80 and HEX80) for the coincident detection of multiple electrons and ions, as described for a similar setup by Ablikim \textit{et al.} \cite{Ablikim2019AMolecules}. For the experiment described here, only the ion detector is used. The three-dimensional momentum vectors are obtained from the ions' time of flight and hit positions as described by Lam \textit{et al.}\cite{Lam2024DifferentiatingImaging} and Ablikim \textit{et al.} \cite{Ablikim2019AMolecules}. From these momentum vectors, we determine the KER and the angular correlations between the ionic momenta of all the coincidence channels that are of interest.

\subsection{Coulomb explosion simulation}

To guide the search for a possible isomer and identify the different reaction channels, we performed classical Coulomb explosion simulations for the molecular geometries corresponding to the $\mathrm{CH_{2}I_{2}}$ equilibrium geometry, the iso-$\mathrm{CH_{2}I_{2}}$, and the C--I bond cleavage pathway including rotation of the $\mathrm{CH_{2}I}$ intermediate. The simulations are based on solving Newton’s equations of motion for three point charges located at the positions of the carbon and iodine atoms. To match the experimental observables, the $\mathrm{CH_{2}}$ fragment was kept intact. It is treated as a point particle in the simulations. 
Furthermore, the ionization and fragmentation process induced by the probe pulse was assumed to be instantaneous (i.e., instantaneously breaking both C--I bonds and resulting in a charge q=1 on the carbon atom and q=2 on each of the two iodine atoms), and the repulsion between the three fragments was treated as purely Coulombic. From prior experience, the calculations based on these assumptions typically overestimate the resulting KER but reproduce the experimentally observed momentum correlations \cite{Lezius2002PolyatomicDynamics, Legare2005LaserMolecules, Liu2007IonizationField, Corrales2012VelocityPulses, Ablikim2016IdentificationImaging, Pathak2020DifferentiatingImaging, Bhattacharyya2022Strong-Field-InducedTribromomethane, Lam2024DifferentiatingImaging}.

The optimized structures of the molecule in the equilibrium geometry of the neutral electronic ground state and the isomer geometry were taken from Borin \textit{et al.} \cite{Borin2016DirectStudy}. 
For each case, an ensemble of 5000 geometries was generated by randomly varying the initial atomic positions within a radius of 0.1~\AA\space around the equilibrium geometries and the initial velocities by up to $\mathrm{2x10^{-4}}$~a.u., respectively. 
These parameters were chosen empirically in order to approximately match the width of the fragment kinetic energy and momentum distributions observed in the experiment. 
In the simulation for the dissociation channel, a rotational period of \(\sim\)300~fs (obtained experimentally for the two-body dissociation into $\mathrm{CH_{2}I}$ + I following UV absorption and C--I cleavage, see Fig.~S11) was used for the $\mathrm{CH_{2}I}$ intermediate, and the translational KEs used for the ion pairs were obtained from the experimental asymptotic KER for the respective channels. 

\section{\label{sec:Res&Dis}Results and discussions}
\begin{figure*}
\includegraphics[width=\textwidth]{Figures_290/975_3body_NewtonMaps.png}
\caption{\label{fig:IRNewtonMaps}\blue{Newton plots of the (a) $\mathrm{CH_{2}^{+} + I^{+} + I^{+}}$, (b) $\mathrm{CH_{2}^{+} + I^{2+} + I^{+}}$, and (c) $\mathrm{CH_{2}^{+} + I^{2+} + I^{2+}}$ coincidence channels for ionization by the NIR pulse alone. 
The momenta are normalized to the magnitude of the $\mathrm{CH_{2}^{+}}$ fragment, and the reference frame is chosen such that the $\mathrm{CH_{2}^{+}}$ momentum is aligned along the $x$-axis, with the $xy$ plane defined by the first detected iodine ion in each channel.
The circular features in (a) and (b) are clear indications of sequential breakup.}}
\end{figure*}

\begin{figure}
\includegraphics[width=\columnwidth]{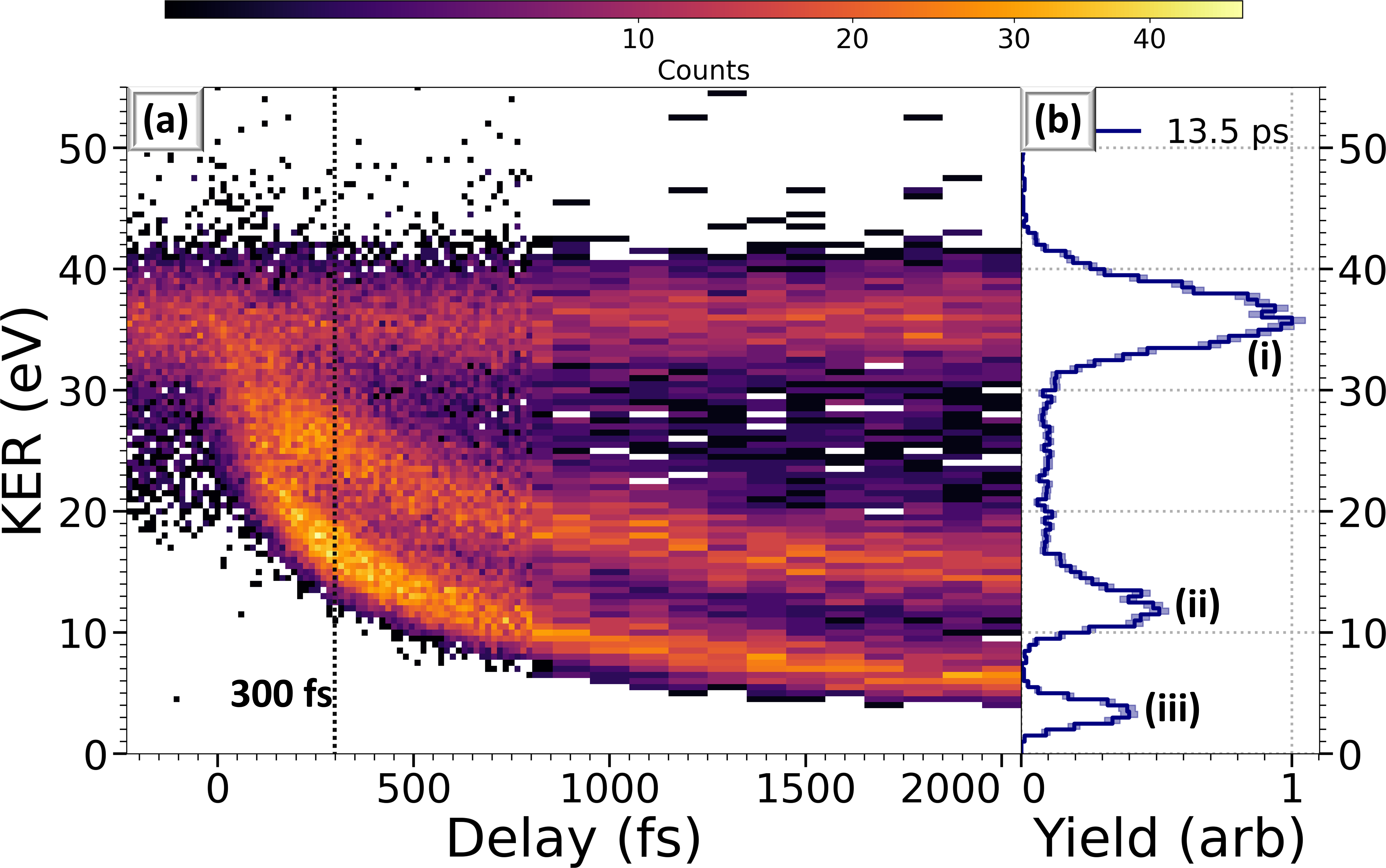}
\caption{\label{fig:delayKER_width}(a) $\mathrm{CH_{2}^{+} + I^{2+} + I^{2+}}$ ion coincidence yield as a function of pump-probe delay and KER. Positive delays correspond to the NIR probe pulse arriving after the UV pump pulse. 
A step size of 15 fs and 100 fs was used for delays up to and beyond 800 fs, respectively.  (b) KER spectra of the same coincidence channel recorded at a fixed delay of 13.5~ps. 
The data shown here were recorded at a pump wavelength of 290 nm. 
The equivalent plots for this and all the following figures with the data recorded at 330 nm are shown in Section II of the SM. 
The vertical dashed line at 300~fs in panel~(a) indicates the experimentally obtained rotational period of the $\mathrm{CH_{2}I}$ fragment.}
\end{figure}

While the main objective of this work is to investigate the transient photoisomerization of diiodomethane ($\mathrm{CH_{2}I_{2}}$) after UV photoabsorption, the majority of the photoexcited molecules undergo direct dissociation following C--I bond cleavage, and the first task is therefore to identify the signatures of these competing reaction pathways. The intense NIR probe pulse ionizes both, the unpumped and the photoexcited molecules, resulting in multiple channels with different final charge states and different ionic fragments
(see Figs.~S1-~S3 of the Supplementary Material (SM) for ion time-of-flight (ToF) mass spectra and multi-ion coincidence spectra). 
 Figure~\ref{fig:IRNewtonMaps} shows the Newton plots for the three dominant three-body fragmentation channels involving emission of two iodine fragments: CH$_{2}^{+}$ + I$^{+}$ + I$^{+}$, CH$_{2}^{+}$ + I$^{2+}$ + I$^{+}$ and CH$_{2}^{+}$ + I$^{2+}$ + I$^{2+}$. 
The former two exhibit strong contributions from sequential breakup, as evident from the circular features (Figs.~\ref{fig:IRNewtonMaps}(a) \& (b)) associated with formation and rotation of an intermediate CH$_{2}$I fragment.
In contrast, the CH$_{2}^{+}$ + I$^{2+}$ + I$^{2+}$ channel in Fig.~\ref{fig:IRNewtonMaps}(c) displays a purely concerted breakup without significant contribution from sequential breakup. 
In the following, we have therefore selected the CH$_{2}^{+}$ + I$^{2+}$ + I$^{2+}$ channel as the most suitable probe of the UV-induced dynamics, since sequential fragmentation would obscure the identification of weaker pathways and complicate the analysis (see Fig.~S4 of the SM).  
In Fig.~\ref{fig:delayKER_width}(a), the ion yield of this coincident channel is plotted as a function of the KER and the pump-probe delay. 

\begin{figure*}
\includegraphics[width=\textwidth]{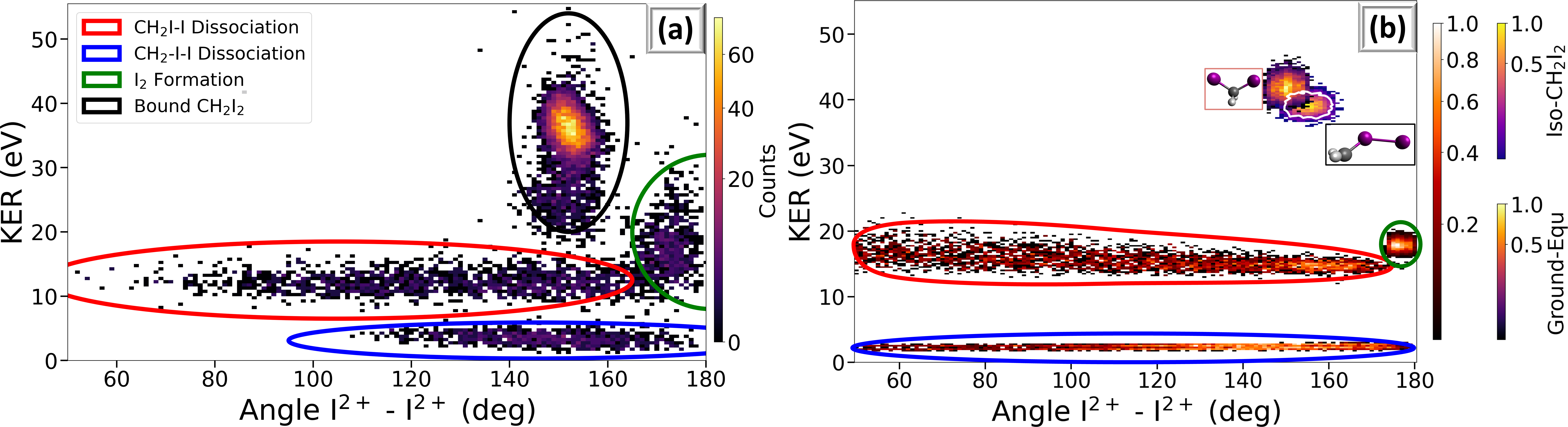}
\caption{\label{fig:KERangAsym}  (a) $\mathrm{CH_{2}^{+} + I^{2+} + I^{2+}}$ ion coincidence yield as a function of KER and angle between the momentum vectors of the two iodine dications for a fixed pump-probe delay of 13.5~ps. Four distinct regions are marked by colored ovals corresponding to different contributions to the coincidence ion yield: Black: Coulomb explosion of bound molecules in or near the equilibrium geometry. Red: C--I cleavage and dissociation to $\mathrm{CH_{2}I}$ + $\mathrm{I}$. Green: Molecular iodine ($\mathrm{I_2}$) formation after UV absorption. Blue: UV-induced three-body dissociation into $\mathrm{CH_{2}}$ + I + I. (b) Coulomb explosion simulations of $\mathrm{CH_{2}I_{2}}$ for the same $\mathrm{CH_{2}^{+} + I^{2+} + I^{2+}}$ channel observed in the experiment, for different molecular geometries: the ground-state equilibrium geometry, the iso-$\mathrm{CH_{2}I_{2}}$ geometry according to Borin \textit{et al.} \cite{Borin2016DirectStudy}, and  for the three dissociation processes that are identified and marked in (a). 
The white contour line in (b) marks the region containing 90\% of the simulated iso-$\mathrm{CH_{2}I_{2}}$ events, showing its partial overlap with the simulated events from the ground-state equilibrium geometries.}
\end{figure*}

Three distinct features are visible in this plot: (i) a horizontal band in the KER range between 30 and 40 eV, which corresponds to the Coulomb explosion of bound (i.e.~non-dissociating) molecules that may or may not have absorbed a UV photon and that were five-fold ionized by the probe pulse; (ii) a curved feature whose KER decreases with increasing pump-probe delay and which emerges from the horizontal band near time-zero, reaching a KER of approximately 15--18~eV at a delay of 2~ps and 9--14~eV at a delay of 13.5~ps (see Fig.~\ref{fig:delayKER_width}(b)); and (iii) another curved feature whose KER decreases even faster than that of feature (ii), reaching a KER of approximately 8--10~eV at a delay of 2~ps and 1--5~eV at a delay of 13.5~ps, at which point the KER no longer changes with delay. Features (ii) and (iii) originate from the molecules that dissociated upon UV photoabsorption, with feature (ii) corresponding to two-body dissociations into $\mathrm{CH_{2}I}$ and $\mathrm{I}$ or $\mathrm{CH_{2}}$ and $\mathrm{I_{2}}$, while feature (iii) corresponds to direct three-body dissociation by the UV pulse into $\mathrm{CH_{2}}$, I and I. While these assignments can be made with good confidence based on the asymptotic KERs of both channels, further insights and confirmation are obtained when inspecting the angular correlation between the momentum vectors of the ionic fragments\cite{KaderiyaImagingSpectroscopy}. 

Figure~\ref{fig:KERangAsym}(a) shows the coincidence ion yield at the asymptotic delay of 13.5~ps as a function of the KER and the angle between the momentum vectors of two iodine dications. The most intense feature at a KER of 30--40~eV (with a smaller side-peak between 20--30~eV) and relatively well defined ($\mathrm{I^{2+}}$, $\mathrm{I^{2+}}$) angle centered around 150$^\circ$ (marked by the black ellipse in Fig.~\ref{fig:KERangAsym}(a)) stems from the Coulomb explosion of bound molecules in or near their equilibrium geometry, as confirmed by the data taken without the UV pulse present (see Fig.~S4 in the SM). At a KER of approximately 15~eV (red ellipse), an angularly broad feature spanning 60--180 degrees in the ($\mathrm{I^{2+}}$, $\mathrm{I^{2+}}$) angle can be attributed to C--I cleavage and dissociation into $\mathrm{CH_{2}I}$ + $\mathrm{I}$ induced by a single-UV-photon absorption. The broad spread in angle results from high rotational excitation of the $\mathrm{CH_{2}I}$ fragment due to the torque imparted from the C--I bond cleavage, as also observed in other dihalomethanes \cite{Murillo-Sanchez2018Halogen-atomBrI, Allum2018CoulombDynamics, MarggiPoullain2018PhotodissociationStudy, Kockert2022UV-inducedPulses, Recio2022ImagingNm, Walmsley2023CharacterizingImaging}.

Two additional features, each of them with approximately half the number of events as in the red ellipse discussed above, are prominent in Fig.~\ref{fig:KERangAsym}. First, a rather localized spot peaked at a KER slightly below 20 eV and an ($\mathrm{I^{2+}}$, $\mathrm{I^{2+}}$) angle close to 180° (green ellipse) can be uniquely attributed to $\mathrm{I_{2}}$ formation after UV excitation (i.e., dissociation into $\mathrm{CH_{2}}$ + $\mathrm{I_{2}}$)
since both the KER and the back-to-back emission of the two iodine ions are consistent with a Coulomb explosion of $\mathrm{I_{2}}$. 
Second, a clearly separated contribution at low KER (blue ellipse) is attributed to the UV-induced three-body dissociation into $\mathrm{CH_{2}}$ + I + I. Both of these features result mainly from multi-photon excitation by the UV pulse. 
This assignment is based on (i) the energy required to trigger three-body dissociation (4.8 - 5 eV \cite{Chen2011I2Spectroscopy,Toulson2016Near-UVCH2I2}), (ii) earlier experimental work that reported rather low quantum yield of $\mathrm{I_{2}}$ elimination upon single-photon excitation \cite{Chen2011I2Spectroscopy}, and (iii) on the observed dependence on the UV power. 
The log--log analysis, as shown in Fig.~S15, reveals that the C--I dissociation yield scales linearly with pump intensity (slope $\approx 1$), consistent with a one-photon absorption process.  
In contrast, the molecular I$_{2}$ formation and the CH$_{2}$ + I + I three-body fragmentation scale nonlinearly (slope $\approx 2$), confirming their two-photon character.
A detailed analysis of the UV power dependence of the different channels for pump wavelengths of 290 nm and 330 nm is presented in Sec.~IV of the SM.

The results of the classical Coulomb explosion simulations for the neutral ground-state geometry and the iso-$\mathrm{CH_{2}I_{2}}$ equilibrium geometry are shown in Fig.~\ref{fig:KERangAsym}(b), along with the simulation results for the $\mathrm{CH_{2}I}$ + I, $\mathrm{CH_{2}}$ + I${_{2}}$ and $\mathrm{CH_{2}}$ + I + I dissociation channels at 13.5~ps pump-probe delay. Apart from overestimating the measured KER, most likely due to the assumption of a purely Coulombic potential, 
the simulated results qualitatively match the experimental observations for all channels, validating the use of these simulations in identifying the observables corresponding to the possible formation of the isomer.

Although the simulations show a subtle difference in the KER of the equilibrium geometry and the isomer, this distinction can be expected to be much less pronounced in the experiment, where a significant spread in the KER is observed even when no UV pulse is present. However, the simulations also predict that the Coulomb explosion of the isomer will lead to some ion yield at larger angles between the ($\mathrm{I^{2+}}$, $\mathrm{I^{2+}}$) momenta than those realized by the Coulomb explosion of the equilibrium geometry. In the experiment, essentially no events are observed with ($\mathrm{I^{2+}}$, $\mathrm{I^{2+}}$) angles larger than 160$^\circ$ when no UV pulse is present (see Fig.~S4 in the SM), but a transient signal with high KER is observed in this region for pump probe delays between 0 and 250 fs, see movie in Fig.~S16 (Multimedia available online). By selecting only those events in the experimental data where the ($\mathrm{I^{2+}}$, $\mathrm{I^{2+}}$) angle is larger than 160$^\circ$, we can thus almost completely suppress the contribution from molecules in the equilibrium geometry.

\begin{figure}
\includegraphics[width=\columnwidth]{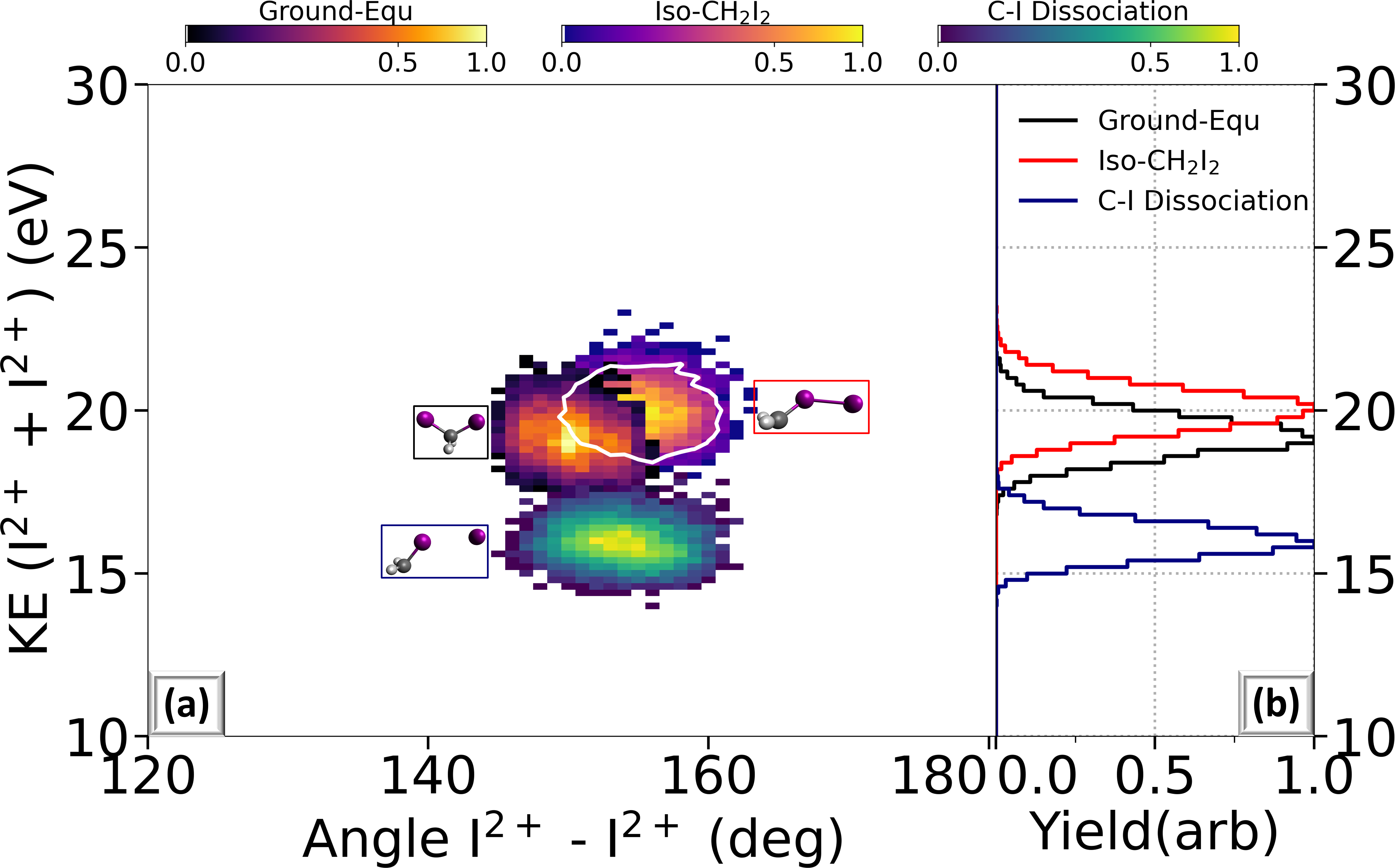}
\caption{\label{fig:SimKESumAng}\blue{Coulomb explosion simulations of $\mathrm{CH_{2}I_{2}}$ for the $\mathrm{CH_{2}^{+} + I^{2+} + I^{2+}}$ breakup channel for different molecular geometries: the ground-state equilibrium geometry, the iso-$\mathrm{CH_{2}I_{2}}$ geometry according to Borin \textit{et al.} \cite{Borin2016DirectStudy}, and for a specific geometry during the C--I dissociation process when the $\mathrm{CH_{2}-I-I}$ angle in the dissociating molecule is the same as in iso-$\mathrm{CH_{2}I_{2}}$ due to the rotation of the $\mathrm{CH_{2}I}$ radical. 
The two panels show the distribution of these events (a) as a function of the KE sum of the two iodine fragments and angle between the two $\mathrm{I^{2+}}$ momentum vectors, and (b) normalized yield of the KE sum of the two $\mathrm{I^{2+}}$ ions integrated for all angles. 
To emphasize the structure associated with iso-$\mathrm{CH_{2}I_{2}}$ and its overlap with the other distributions, a white contour representing the region containing 90\% of the iso-$\mathrm{CH_{2}I_{2}}$ event density is overlaid on panel (a).}}
\end{figure}

However, from the experimental data, it is evident that the three dissociation pathways---(i) $\mathrm{CH_{2}I}$-I two-body dissociation with $\mathrm{CH_{2}I}$ rotation, (ii) $\mathrm{CH_{2}}$-I-I three-body dissociation and (iii) molecular $\mathrm{I_{2}}$ formation---could also contribute to the region with the ($\mathrm{I^{2+}}$, $\mathrm{I^{2+}}$) angle larger than 160$^\circ$ (at least at large delays), as seen in Fig.~\ref{fig:KERangAsym}(a). Contributions from these pathways need to be filtered out to identify possible signatures of isomerization to $\mathrm{CH_{2}I}$--I. The simulations in Fig.~\ref{fig:SimKESumAng} show that after Coulomb explosion, the sum of the KEs of the two $\mathrm{I^{2+}}$ fragments is larger for the iso-$\mathrm{CH_{2}I_{2}}$ molecules than that of the $\mathrm{CH_{2}I_{2}}$ molecules that undergo direct dissociation into $\mathrm{CH_{2}I}$ and I via C--I cleavage. Furthermore, in the cases of three-body dissociation or molecular iodine formation, the KE of the methyl ion decreases rapidly with pump-probe delay as the $\mathrm{CH_{2}}$ fragment quickly moves away from the two iodine atoms or the $\mathrm{I_{2}}$ molecule. Therefore, the sum of the KEs of the two $\mathrm{I^{2+}}$ fragments as well as the KE of the $\mathrm{CH_{2}}$ fragment can serve as additional parameters to discriminate the different channels.

Since it is very likely that any isomer-like geometries, were they to be formed as predicted by Borin \textit{et al.}, would be visited quickly due to the large amount of internal energy deposited into the molecule by the UV photon, we will concentrate our search on the smaller delays below 1~ps, where the isomer-geometry should appear as an additional contribution in the region corresponding to the bound molecules. Note that we have, nonetheless, also searched at larger pump-probe delays but have not found any statistically significant indications of isomer-like geometries at larger delays.

\begin{figure}
\includegraphics[width=\columnwidth]{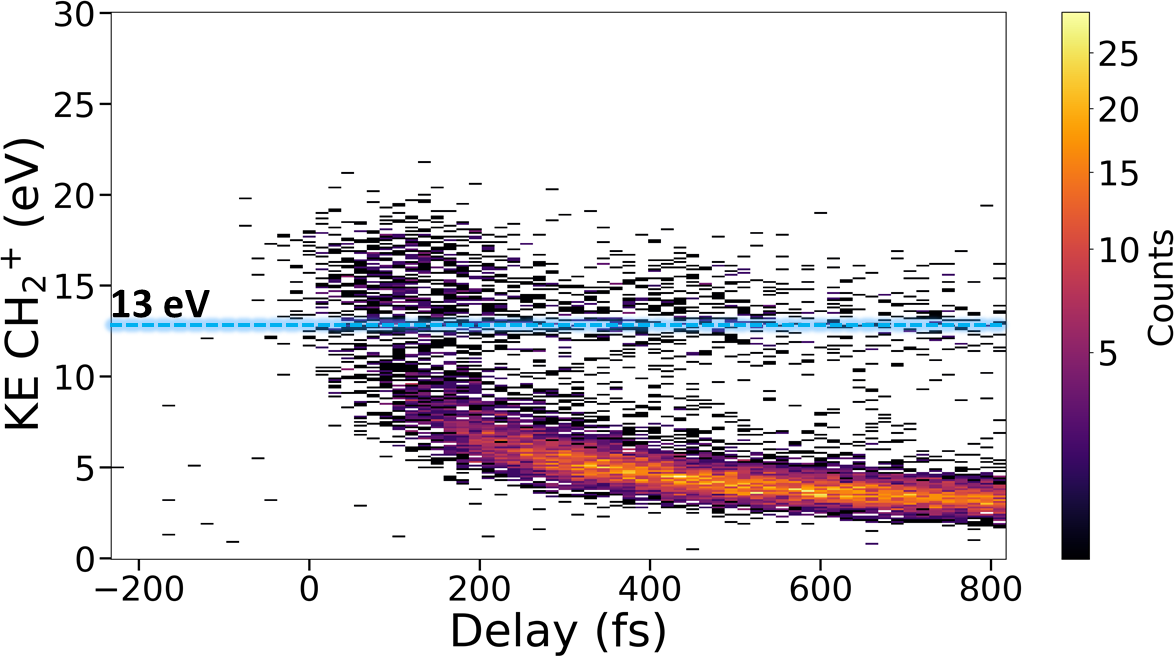}
\caption{\label{fig:KE1delayLargeAngle}Kinetic energy of the $\mathrm{CH_{2}^{+}}$ fragment as a function of the pump-probe delay, with only those coincidence events shown where the angle between the two $\mathrm{I^{2+}}$ momentum vectors is greater than 160$^\circ$. The region below the blue line, at 13 eV kinetic energy of methyl cation, corresponds to events from $\mathrm{CH_{2}}$-I-I three-body dissociation and molecular $\mathrm{I_{2}}$ formation pathways (see text). The region of interest, above 13 eV, is where any events from photoisomerization following UV excitation would appear, along with the contribution from two-body dissociation producing I (or $\mathrm{I^{*}}$) and rotationally excited $\mathrm{CH_{2}I}$. 
}
\end{figure}

In Fig.~\ref{fig:KE1delayLargeAngle}, the KE distribution of the methyl cation ($\mathrm{CH_{2}^{+}}$) is shown as a function of the pump-probe delay up to 800~fs and only for those events where the ($\mathrm{I^{2+}}$, $\mathrm{I^{2+}}$) angle is larger than 160$^\circ$. 
Based on the reasoning outlined in the previous paragraph, we can conclude that the events with quickly decreasing KE, below the blue line in Fig.~\ref{fig:KE1delayLargeAngle}, correspond to either $\mathrm{CH_{2}}$-I-I three-body dissociation or molecular iodine formation. 
The events above the blue line at 13~eV mainly correspond to dissociation of $\mathrm{CH_{2}I_{2}}$ into $\mathrm{CH_{2}I}$ and $\mathrm{I}$ (with subsequent rotation of the $\mathrm{CH_{2}I}$ fragment) as well as any possible formation of iso-$\mathrm{CH_{2}I_{2}}$, with only spurious contributions from the three-body dissociation and molecular iodine formation pathways at small delays below 200~fs.
The choice of 13~eV as the boundary is motivated by the experimental data (Fig.~S5), where a distinct separation is observed between the  decaying low-KE band and the high-KE island in the 100--200~fs region. 
Although the constant 13~eV horizontal line intersects other features at later delays, our analysis focuses only on the early-delay (<~300~fs) region, so those intersections do not affect the conclusions.

\begin{figure}
\includegraphics[width=\columnwidth]{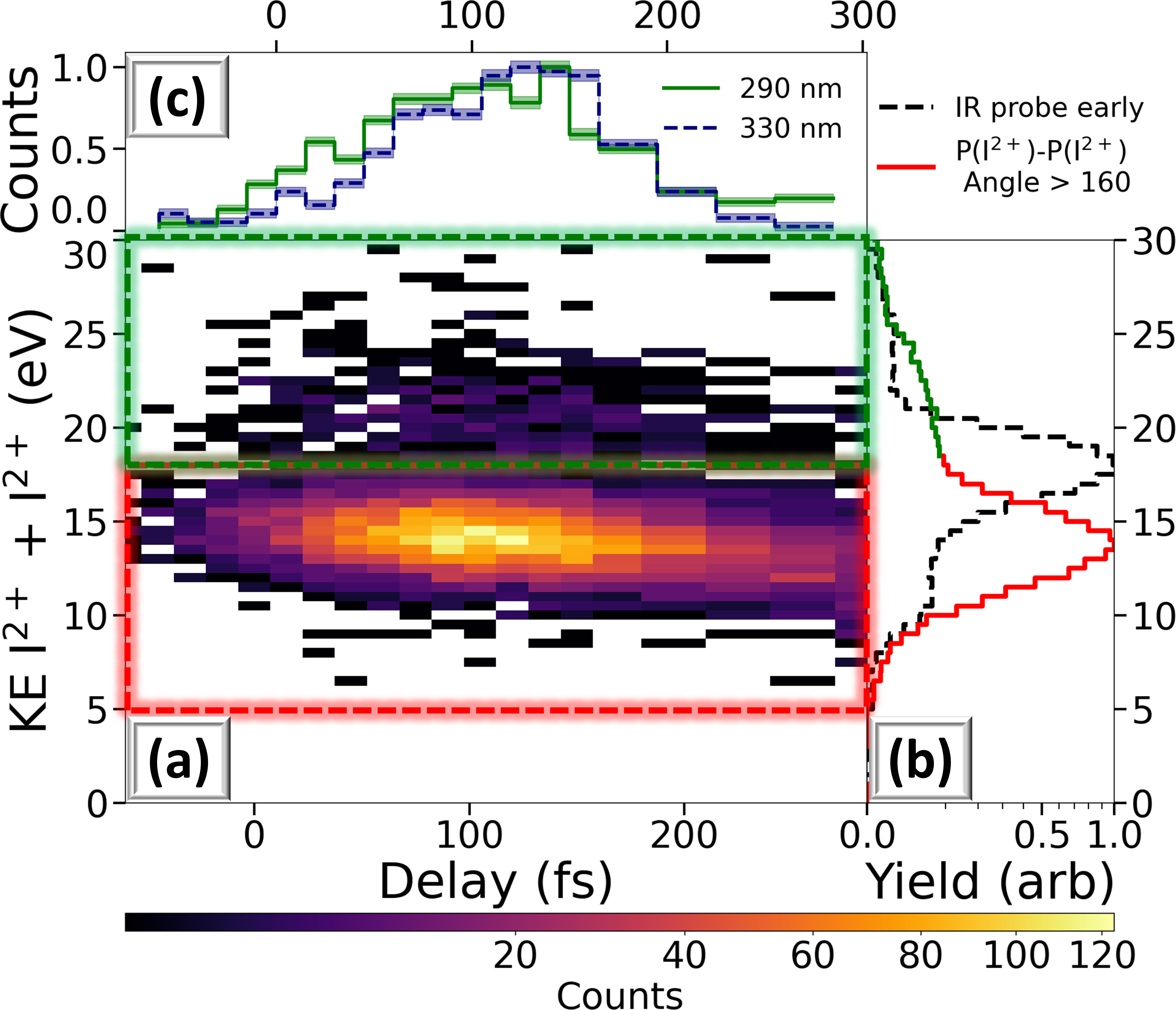}
\caption{\label{fig:KE23SumSrtRng} (a) Coincidence ion yield as a function of pump-probe delay and KE sum of the two $\mathrm{I^{2+}}$ fragments, for only those events where the angle between the two $\mathrm{I^{2+}}$ momentum vectors is greater than 160° and the KE of the $\mathrm{CH_{2}^{+}}$ fragment is greater than 13~eV (above the blue line in Fig.~\ref{fig:KE1delayLargeAngle}). The two regions of interest are described in the text. (b) Projection of the coincident ion yield in panel (a) on the KE axis. The corresponding KE sum distribution for "unpumped" molecules (NIR probe pulse arrives before the UV pump pulse) is shown as a black dotted line.
(c) Projection of the coincident ion yield in the green ROI in (a) on the pump-probe delay axis (green), compared to the similarly analyzed yield at a pump wavelength of 330~nm (blue).}
\end{figure}

Next, we select only those events with kinetic energy of $\mathrm{CH_{2}^{+}}$ ionic fragment above 13~eV from Fig.~\ref{fig:KE1delayLargeAngle} and plot the sum of the KEs of the two iodine fragments as a function of pump-probe delay in Fig.~\ref{fig:KE23SumSrtRng}(a). The plot reveals two contributions: an intense feature with a KE sum between 10 and 17~eV (marked by the red rectangle), and a much weaker feature at a higher KE sum, marked by the green rectangle, which contains approximately 7\% of the events in the intense feature. 
The projection in panel (b) reveals that these two contributions are shifted toward lower (solid line, red section) and higher energies (solid line, green section), respectively, compared to the KE sum of the unpumped molecules (black dotted line).
The integrated yield inside the green region of interest as a function of the pump-probe delay is shown as green line in Fig.~\ref{fig:KE23SumSrtRng}(c). 
Based on the simulations shown in Fig.~\ref{fig:SimKESumAng}(b), we assign the events with the higher KE sum to molecular geometries resembling the iso-$\mathrm{CH_{2}I_{2}}$ structure predicted by Borin \textit{et al.}, which have a slightly smaller I--I distance than geometries resulting from the dissociation of $\mathrm{CH_{2}I_{2}}$. 
To quantify the change in the I--I distance, we calibrate the relationship between the KE sum of the two I$^{2+}$ ions and the I--I distance using Coulomb explosion simulations with varied I--I distances.  
Applying this calibration to the experimental data yields an estimated I--I distances of $\sim$3.0~Å for the events with high KE sum and $\sim$4.6~Å for the events with low KE sum (see Fig.~S8), compared to $\sim$3.58~Å in the equilibrium geometry.
Fig.~\ref{fig:KE23SumSrtRng}(c) shows that the occurrence of these iso-$\mathrm{CH_{2}I_{2}}$-like structures peaks at a delay of approximately 120~fs and quickly decays again on a similar time scale. 
Analysis of the pump-probe data recorded at a pump wavelength of 330~nm, following the exact same steps as described above (see Sec.~III in SM), yields the blue dashed line in Fig.~\ref{fig:KE23SumSrtRng}(c), which shows the same behavior as the data at 290~nm except for possibly a subtle shift towards slightly longer delays, which is at the borderline of statistical significance for the present data set.
We focus on the 290~nm results (which have superior statistics and signal-to-noise ratio due to the higher absorption cross section at 290~nm\cite{KaderiyaImagingSpectroscopy}) in the main text, while noting that the 330~nm data, which is shown in more detail in the SM, exhibit qualitatively similar dynamics, as can be seen in Fig.~\ref*{fig:KE23SumSrtRng}(c).

\begin{figure*} 
\includegraphics[width=\textwidth]{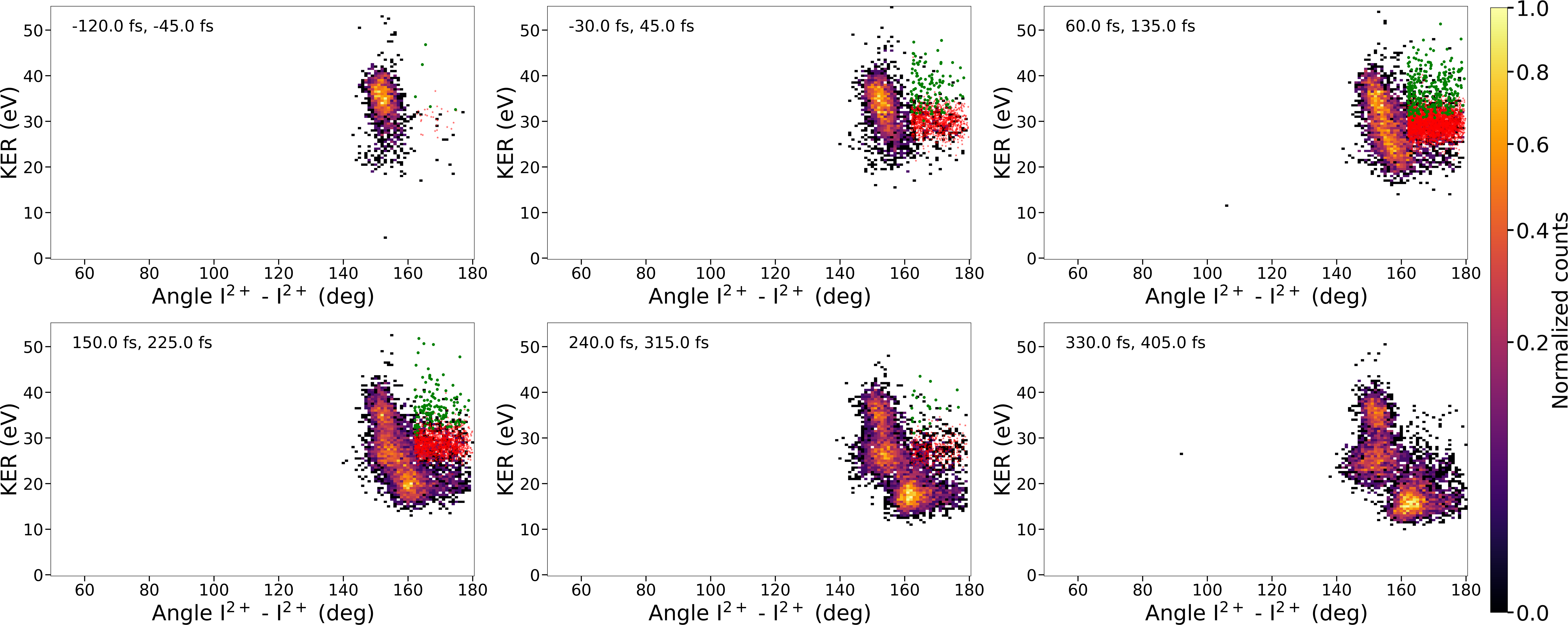}
\caption{\label{fig:delayslicedKERAng}\blue{Delay-sliced snapshots of the CH$_2^+$ + I$^{2+}$ + I$^{2+}$ coincidence channel,  showing the coincident ion yield as a function of the total KER and the angle between the two I$^{2+}$ ion momenta from the time delay windows indicated at the top left of each panel.  
The gated events associated with the two rectangular regions in Fig.~\ref{fig:KE23SumSrtRng} are overlaid as green and red scatter points.}}
\end{figure*}

\begin{figure}
\includegraphics[width=\columnwidth]{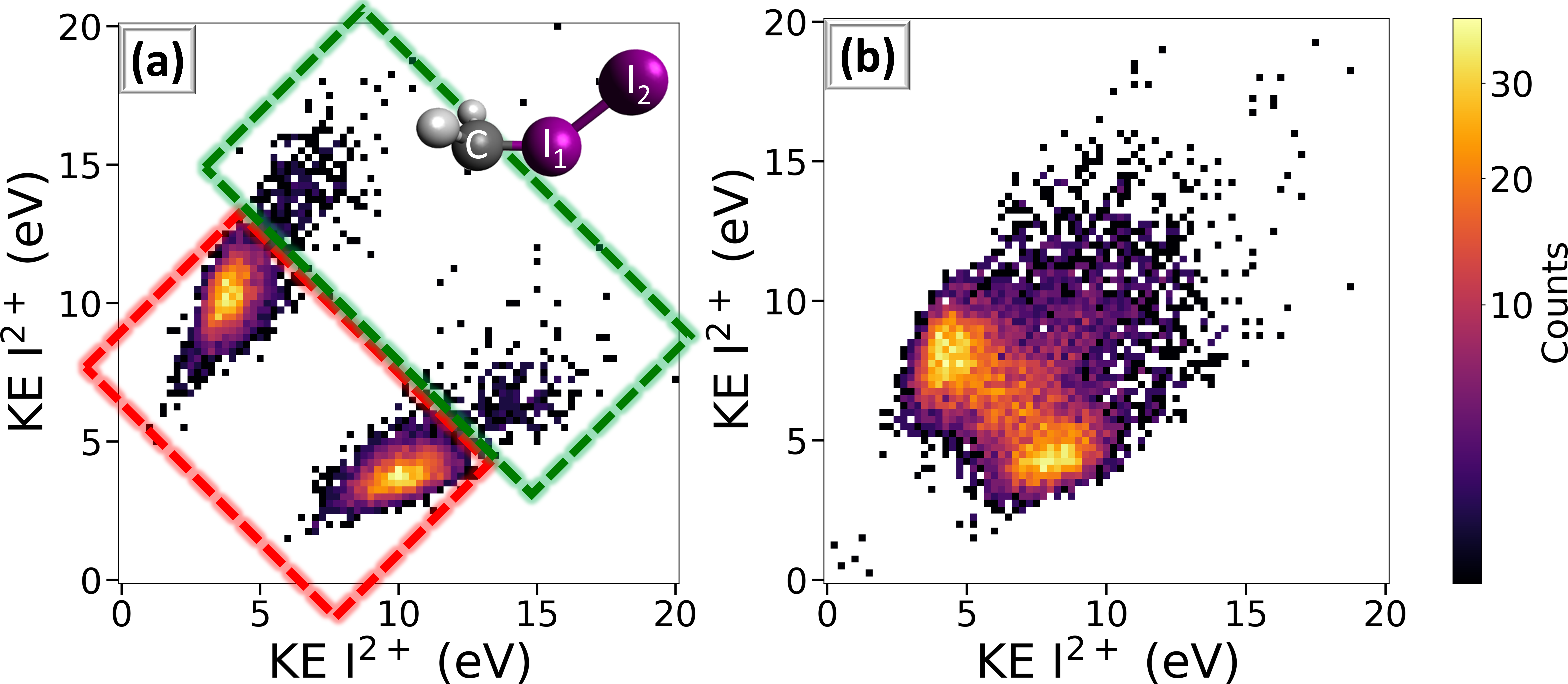}
\caption{\label{fig:KE2vsKE3SrtRng} Energy sharing between the two $\mathrm{I^{2+}}$ fragments, integrated over the pump-probe delays shown in Fig.~\ref{fig:KE23SumSrtRng}, for only those events where the angle between the two $\mathrm{I^{2+}}$ momentum vectors is greater than 160$^\circ$ and (a) the KE of the $\mathrm{CH_{2}^{+}}$ fragment is greater than 13~eV (above the blue line in Fig.~\ref{fig:KE1delayLargeAngle}); or (b) the KE of the $\mathrm{CH_{2}^{+}}$ fragment is less than 13~eV (below the blue line in Fig.~\ref{fig:KE1delayLargeAngle}).
The red and green rectangles in panel (a) correspond to events originating from the respective rectangular regions in Fig.~\ref{fig:KE23SumSrtRng}.
}
\end{figure}

To further demonstrate the time dependence of the iso-CH$_2$I$_2$-like contribution, Fig.~\ref{fig:delayslicedKERAng} presents selected snapshots of the delay-dependent CH$_2^+$ + I$^{2+}$ + I$^{2+}$ coincidence yield as a function of the total KER and the angle between the momenta of the two I$^{2+}$ ions. The events associated with the two rectangular regions in Fig.~\ref{fig:KE23SumSrtRng} are overlaid as green and red scatter points to emphasize the iso-CH$_2$I$_2$-like contributions. The figure shows that this transient feature is confined to a narrow temporal window and disappears at later delays, consistent with the dynamics discussed in Fig.~\ref{fig:KE23SumSrtRng}.  

Additional information on the geometry of the transient species can be obtained by considering the energy sharing between the two detected $\mathrm{I^{2+}}$ fragments, as shown in Fig.~\ref{fig:KE2vsKE3SrtRng}. Symmetric geometries, such as the original $\mathrm{CH_{2}I_{2}}$ geometry, or any molecules excited to symmetric vibrational modes, would lead to equal energy sharing, whereas strongly asymmetric geometries 
such as the predicted iso-$\mathrm{CH_{2}I_{2}}$ species, would lead to a pronounced asymmetry in the energy sharing.

The gated events that survive both filters---large I$^{2+}$--I$^{2+}$ momentum angle and high CH$_2^+$ energy---exhibit distinctly asymmetric energy sharing between the two I$^{2+}$ ions, as shown in  Fig.~\ref{fig:KE2vsKE3SrtRng}(a). 
This is further illustrated in Fig.~S7, where the energy-sharing ratio between the two I$^{2+}$ fragments is shown.
While IR-early events peak sharply near zero, indicating symmetric sharing, the gated events of Fig.~\ref{fig:KE2vsKE3SrtRng}(a) yield a broadened distribution centered near 0.5, confirming their asymmetric character and supporting the assignment to iso-CH$_2$I$_2$-like geometries.
In contrast, Fig.~\ref{fig:KE2vsKE3SrtRng}(b) shows that symmetric geometries dominate when the CH$_2^+$ energy is low, corroborating our assignment of these events to the I$_2$ formation channel. 

At this point, we would like to emphasize that although detailed analysis of the CEI data strongly suggests the transient formation of CH${_{2}}$I--I isomer-like geometries that are consistent with previous literature~\cite{Borin2016DirectStudy}, the mapping from position to fragment momentum space in CEI is not unique~\cite{Sayler2018Nonunique}, and we cannot exclude that other geometries would lead to similar CEI signatures. 
However, we consider such isomer-like geometries to be the most likely reason for our observations, given the geometric constraints imposed by the various observables we considered in conjunction with the observed delay-dependent evolution.
That said, it also needs to be stated that the CEI experiment alone is not able to confirm this as an "isomerization" process, nor can the CEI experiment identify the electronic state of the molecule when it assumes these isomer-like geometries.  

Whether or not the process observed and reported here can or should be referred to as "isomerization" cannot be answered by the experimental data alone
and also depends on the exact definition of "isomerization" one chooses to use. Given the short time that the molecules spend in the isomer-like geometry, which is less than the timescale of the rotation of the $\mathrm{CH_{2}I}$ fragment that is produced by direct C--I bond cleavage, the process might be better described as a \emph{transient passage through isomer-like geometries} rather than an actual trapping in the potential well of the isomer, but these terminologies and interpretations strongly rely on careful theoretical modeling, which is beyond the scope of the present paper.

\section{Conclusion}
We have applied time-resolved three-body Coulomb explosion imaging to study the UV-induced (290~nm and 330~nm) photochemical dynamics in gas-phase $\mathrm{CH_{2}I_{2}}$.
In addition to the primary dissociation into CH$_{2}$I + I via direct cleavage of the C--I bond, the formation of molecular $\mathrm{I_2}$, and the three-body dissociation into CH$_{2}$ + I + I (where the latter two processes are attributed to the absorption of more than one UV photon), we also observe the transient molecular configurations resembling iso-$\mathrm{CH_{2}I_{2}}$ geometries, which are formed within approximately 100~fs after the photoexcitation and completely decay within an additional 100~fs. 
Within the uncertainties of the experimental data, the relative yield, formation and decay time of the transient iso-$\mathrm{CH_{2}I_{2}}$-like products appear to be similar at both excitation wavelengths employed in this study, suggesting that their formation does not depend strongly on the excitation wavelength.
While the present experiment cannot accurately quantify the branching ratio of these transient geometries since the strong-field ionization rate is strongly dependent on the electronic state and molecular geometry, it appears that only a small fraction of excited molecules form isomer-like geometries. 
Therefore, it is conceivable that other ultrafast structural studies, such as the gas-phase UED experiment by Liu \textit {et al.}\cite{Liu2020SpectroscopicDiffraction}, may not have been able to observe these structures since it lacked the required signal-to-noise ratio to discern extremely weak channels and also did not have sufficient temporal resolution to resolve such short-lived transient structures.

\section{Supplementary Material}
Supplementary material includes probe-only spectra, supporting results (Dalitz plots, energy-sharing ratios, and I--I distance estimate), comparison results from 330~nm pump, UV power-dependence measurements, and a movie of the delay evolution.

\begin{acknowledgments}
We gratefully acknowledge the technical staff of the J.R.~Macdonald Laboratory for their excellent support of the experiments; former PhD students Dr.~Seyyed Javad Robatjazi and Dr.~Shashank Pathak for commissioning the double-sided VMI apparatus; and Prof.~Brett Esry for insightful discussions throughout the project. This work was supported by the Chemical Sciences, Geosciences, and Biosciences Division, Office of Basic Energy Sciences, Office of Science, US Department of Energy, grants no.~DE-FG02-86ER13491 and DE-SC0020276 (S.B.), and by the National Science Foundation grant no.~PHYS-2409365 (A.S.V.).
\end{acknowledgments}

\section*{Conflict of Interest}
The authors have no conflicts to disclose.

\bibliography{CH2I2_UV_IR_CEI}
\end{document}



\title{Supplementary Material \\ Imaging transient molecular configurations in UV-excited diiodomethane}
\author{Anbu Selvam Venkatachalam}
\email{anbu@ksu.edu}
\affiliation{James R. Macdonald Laboratory, Physics Department, Kansas State University, Manhattan, KS 66506, USA}
\author{Huynh Van Sa Lam}
\affiliation{James R. Macdonald Laboratory, Physics Department, Kansas State University, Manhattan, KS 66506, USA}
\author{Surjendu Bhattacharyya}
\thanks{Current address: SLAC National Accelerator Laboratory, Menlo Park, CA 94025, USA}
\affiliation{James R. Macdonald Laboratory, Physics Department, Kansas State University, Manhattan, KS 66506, USA}
\author{Balram Kaderiya}
\affiliation{James R. Macdonald Laboratory, Physics Department, Kansas State University, Manhattan, KS 66506, USA}
\author{Enliang Wang}
\thanks{Current address: Department of Modern Physics, University of Science and Technology of China, Anhui, China }
\affiliation{James R. Macdonald Laboratory, Physics Department, Kansas State University, Manhattan, KS 66506, USA}
\author{Yijue Ding}
\thanks{Current address: Department of Chemistry, Southern University of Science and Technology, Shenzhen, Guangdong 518000, China}
\affiliation{James R. Macdonald Laboratory, Physics Department, Kansas State University, Manhattan, KS 66506, USA}
\author{Loren Greenman}
\affiliation{James R. Macdonald Laboratory, Physics Department, Kansas State University, Manhattan, KS 66506, USA}
\author{Artem Rudenko}
\affiliation{James R. Macdonald Laboratory, Physics Department, Kansas State University, Manhattan, KS 66506, USA}
\author{Daniel Rolles}
\email{rolles@ksu.edu}
\affiliation{James R. Macdonald Laboratory, Physics Department, Kansas State University, Manhattan, KS 66506, USA}

\date{\today}

\maketitle

\tableofcontents

\clearpage

\renewcommand{\thefigure}{S\arabic{figure}}

\section{\label{sec:IRPlots} NIR-only spectra (without UV pulse)}

\begin{figure}[h]
\includegraphics[width=0.9\columnwidth]{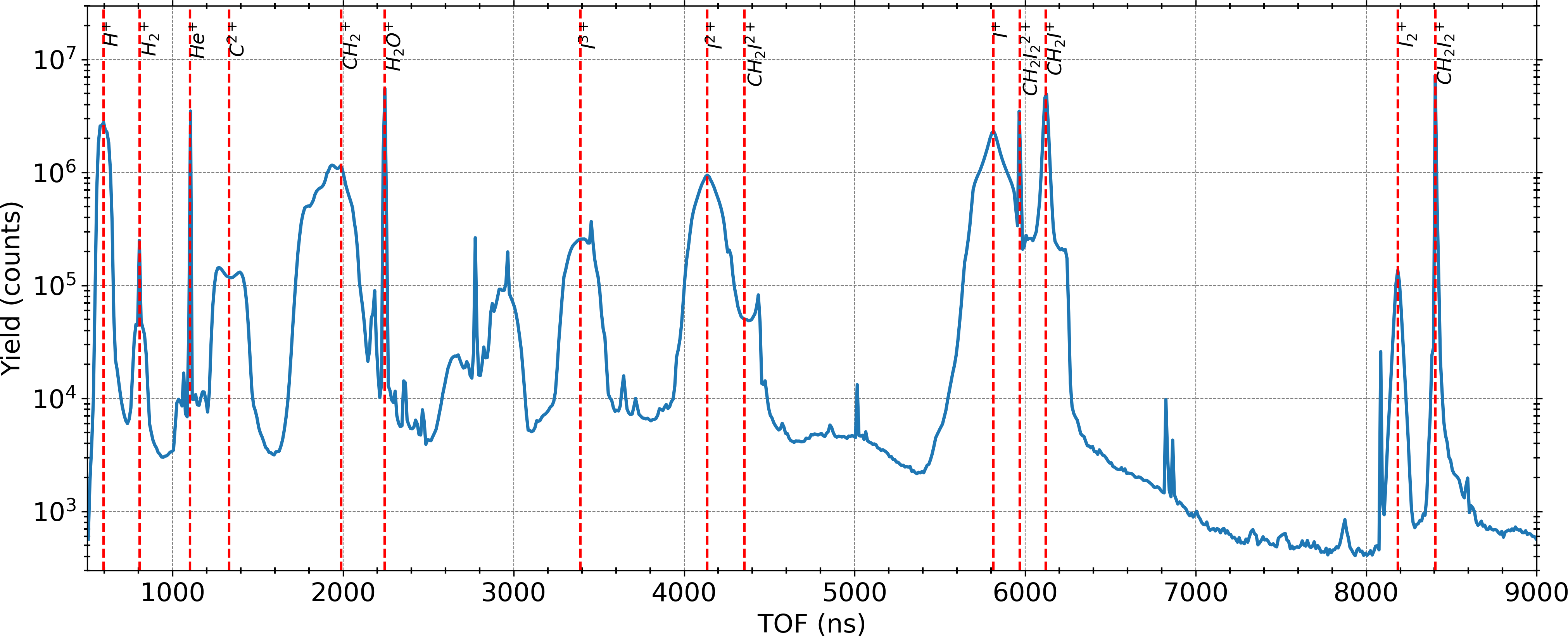}
\caption{\label{fig:SMirToFAll} Ion time-of-flight (ToF) spectrum generated by strong-field ionization of $\mathrm{CH_{2}I_{2}}$ molecule with the NIR pulse only, i.e., without UV pulse. The red vertical lines indicate the calculated ToF for the indicated ions with zero kinetic energy. 
Peaks are labeled based on their respective mass-to-charge ratios.}
\end{figure}

\begin{figure}
\includegraphics[width=0.9\columnwidth]{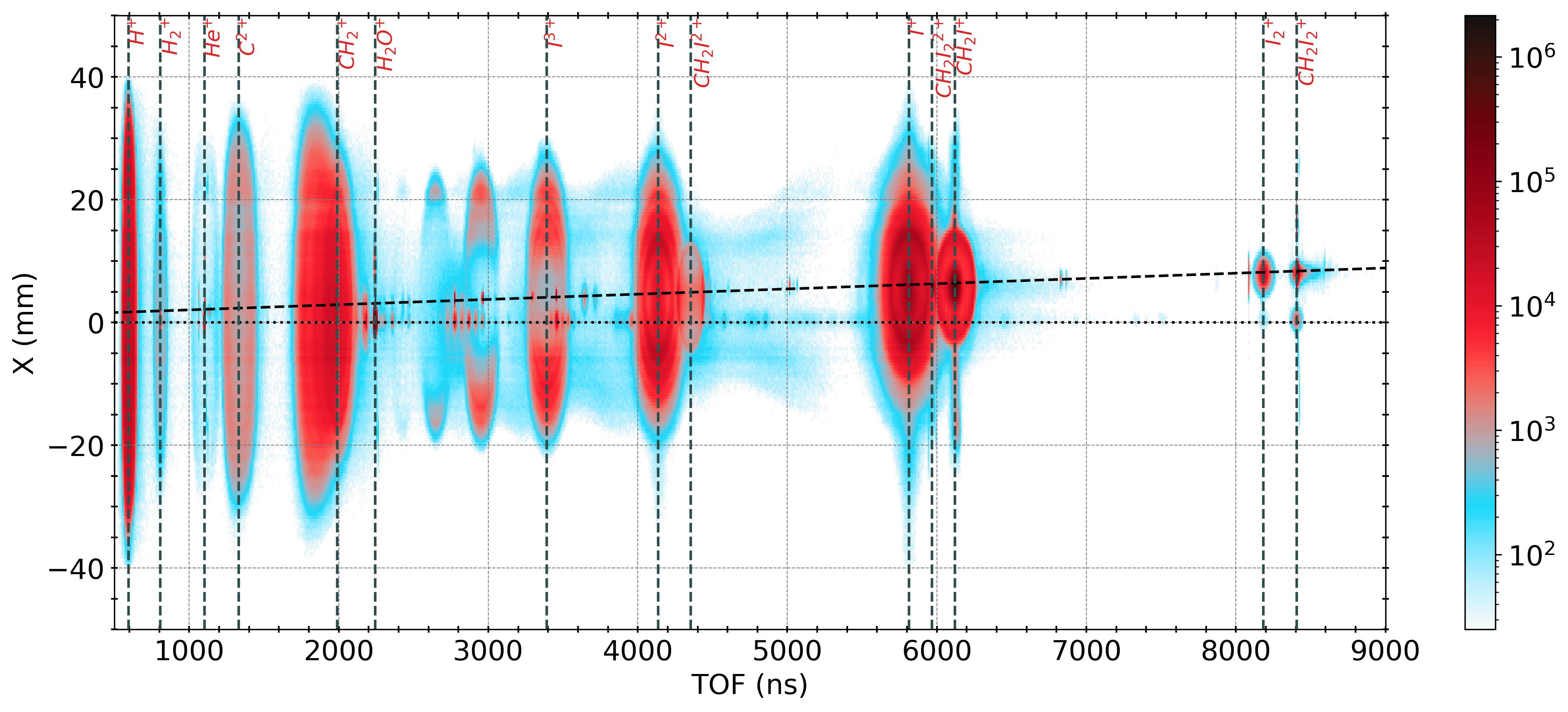}
\caption{\label{fig:SMirToFX} Ion time-of-flight (ToF) spectrum as above but including the detector hit position along the X (molecular beam) direction. The offset in X, indicated by the diagonal dashed line, is due to the velocity of the molecular beam, which helps separate ions generated by the ionization of molecules in the beam from those from residual gas.}
\end{figure}

\begin{figure}
\includegraphics[width=0.8\columnwidth]{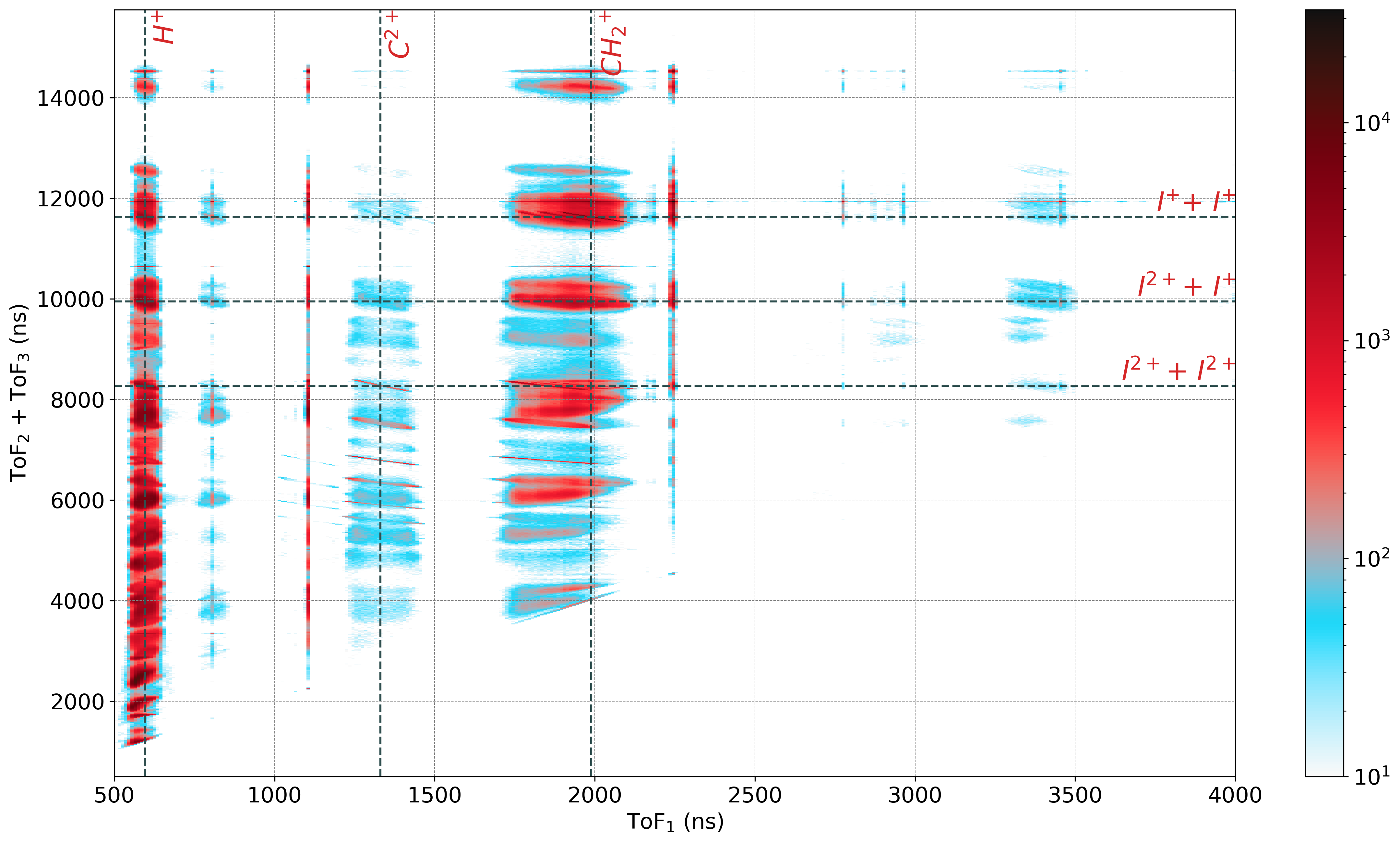}
\caption{\label{fig:SMirTRIPICO} Photoion-photoion-photoion coincidence (TRIPICO) yield generated by strong-field ionization of $\mathrm{CH_{2}I_{2}}$ molecule with the NIR pulse only.
The different coincidence islands correspond to distinct fragmentation channels accessible under probe-only conditions. 
This highlights the various coincident breakup pathways available and provides the basis for identifying the specific probe channel for UV-induced dynamics.}
\end{figure}

\begin{figure}[h]
\includegraphics[width=\columnwidth]{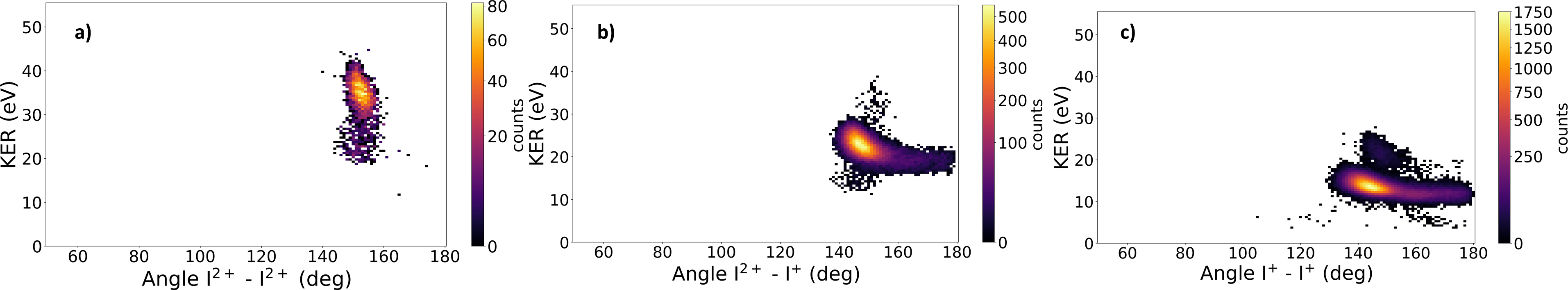}
\caption{\label{fig:SMirKERang23} Ion yield as a function of KER and angle between the momentum vectors of the two iodine ions for (a) $\mathrm{CH_{2}^{+} + I^{+} + I^{+}}$, (b) $\mathrm{CH_{2}^{+} + I^{2+} + I^{+}}$ and (c) $\mathrm{CH_{2}^{+} + I^{2+} + I^{2+}}$ coincidence channels, again for ionization by the NIR pulse alone.
The first two channels exhibit strong sequential breakup contributions, consistent with rotational excitation of the intermediate fragments. 
In contrast, the $\mathrm{CH_{2}^{+} + I^{2+} + I^{2+}}$ channel shows minimal sequential character, making it the most suitable fragmentation channel for probing pump-induced dynamics.}
\end{figure}

\blue{This section presents the spectra obtained using the strong-field NIR probe pulse alone, without UV excitation, in order to establish the probe-only contributions. 
Figure~\ref{fig:SMirToFAll} shows the ion time-of-flight (ToF) spectrum with peaks corresponding to different ionic fragments produced by the strong IR field. 
Figure~\ref{fig:SMirToFX} shows the detector hit position along the molecular beam axis, where the observed offset shows the velocity of the molecular jet and helps distinguish target ions from background contributions. 

Figure~\ref{fig:SMirTRIPICO} displays the three-ion coincidence (TRIPICO) spectrum under probe-only conditions. This highlights the various accessible coincidence channels generated by the NIR pulse and shows the different fragmentation pathways that can be probed with the strong IR field. 
Figure~\ref{fig:SMirKERang23} shows the KER versus I--I momentum angle distributions for three representative coincidence channels with three-, four-, and five-fold total final charge states. 
We focus on the five-fold channel (CH$_2^+$ + I$^{2+}$ + I$^{2+}$), which exhibits minimal contributions from sequential breakup. 
By contrast, the three- and four-fold channels contain significant sequential contributions, as also evident in the Newton maps (Fig.~\ref*{fig:IRNewtonMaps} in the main text). 
Sequential fragmentation introduces rotational excitation of the intermediate fragments, which obscures the signatures of pump-induced dynamics and complicates the identification of weaker competing pathways. 
Thus, the probe-only analysis justifies our channel selection and establishes the baseline for interpreting the UV-pump/IR-probe results.}

\clearpage 
\section{\label{sec:SimExpMorePlots} More simulation and experimental results}

\begin{figure}[h]
\includegraphics[width=0.8\columnwidth]{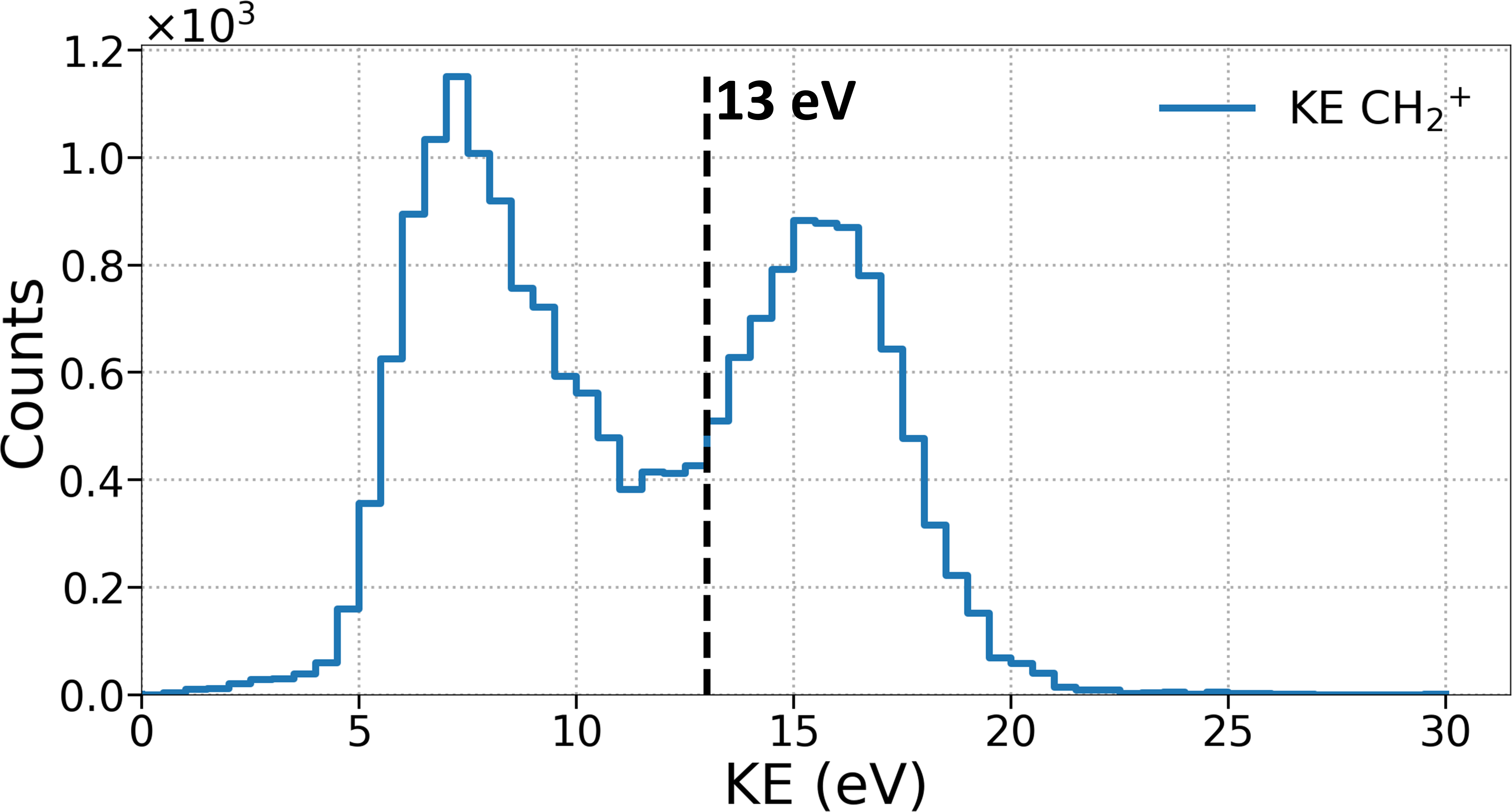}
\caption{\label{fig:SMKE1Gating} \blue{Delay-integrated kinetic energy distribution of the $\mathrm{CH_{2}^{+}}$ fragment from the CH$_{2}^{+}$ + I$^{2+}$ + I$^{2+}$ coincidence channel, gated on $\mathrm{I^{2+}}$--$\mathrm{I^{2+}}$ momentum angles larger than $160^{\circ}$.  
The distribution shows a dip that separates the decaying low-KE band from the higher-KE island observed in the 100--200~fs region (see Fig.~\ref*{fig:KE1delayLargeAngle}).  
The dashed line at 13~eV corresponds to the upper edge of this dip and is used as the threshold to distinguish between these two contributions.}
}
\end{figure}

\begin{figure}[h]
\includegraphics[width=\columnwidth]{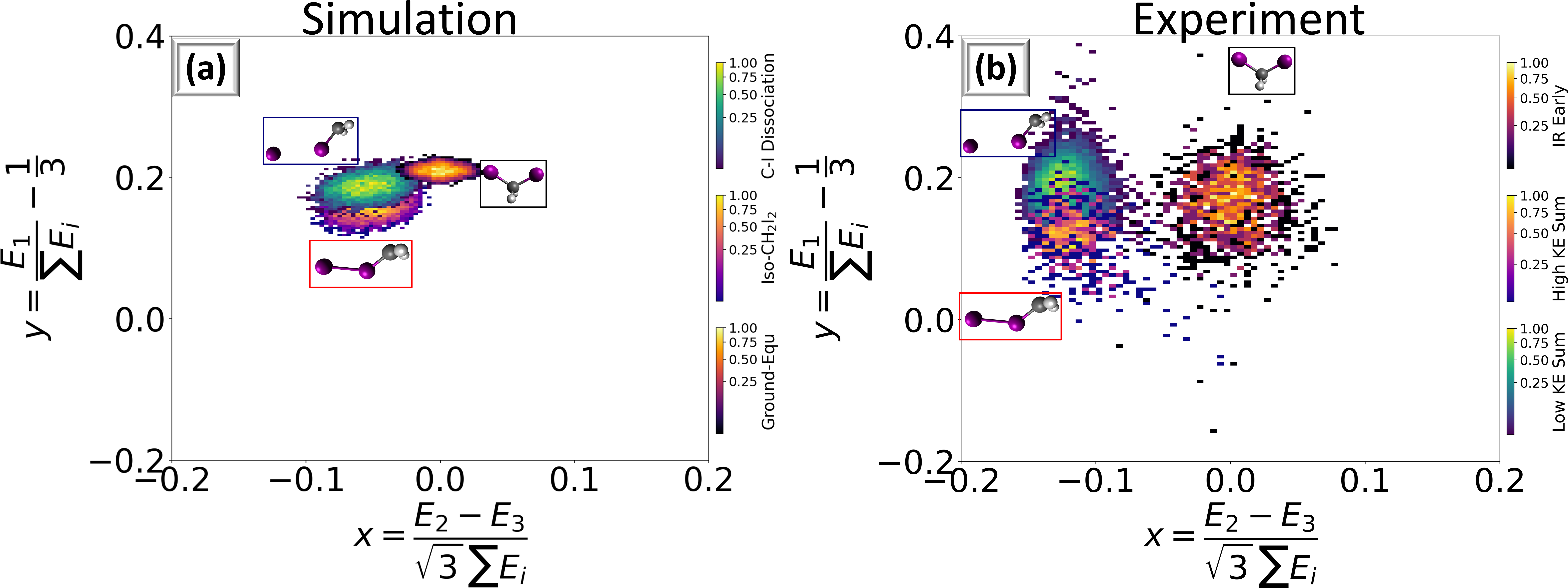}
\caption{\label{fig:SMDalitzPlotExpSim} \blue{Dalitz plots for the $\mathrm{CH_{2}^{+} + I^{2+} + I^{2+}}$ channel from simulation (a) and experiment (b). 
The simulation shows contributions from the ground state, the CH$_{2}$I--I isomer geometry, and one specific geometry along the C--I dissociation pathway where the angle $\angle$(CH$_{2}$--I--I) is constrained to match that of the isomer. 
In the Dalitz representation, symmetric geometries such as the ground-state equilibrium CH$_2$I$_2$ appear centered along the horizontal axis, pointing to equal momentum sharing between the two I$^{2+}$ fragments.
In contrast, the simulated iso-CH$_2$I$_2$ geometry and the C–I dissociation geometry islands are off-center, indicating asymmetric iodine configurations.
Comparison of the two off-center clusters shows that the CH$_{2}^{+}$ fragment carries less fractional energy in the isomer geometry than in the C–I dissociation geometry, consistent with a shorter I--I distance in the iso-CH$_2$I$_2$ configuration.
Panel (b) shows the experimental Dalitz plots corresponding to the same molecular configurations: NIR-early events, which result from the CE of the unperturbed, ground-state geometry, and the two gated regions marked by the red and green boxes in Fig.~\ref*{fig:KE23SumSrtRng}. 
As discussed in the main text, these two regions, which separate high- and low-kinetic-energy sums of the two iodine ions for events with a large I--I momentum-space angle and high CH$_2^{+}$ kinetic energy, correspond to the experimental selection of the contributions from the isomer-like configuration and the C--I dissociation events resulting in similar I--I momentum-space angles, respectively.  
For the high- and low-KE sum events in panel (b), the iodine fragments were sorted by kinetic energy to enable direct comparison with the simulations.
The experimental data exhibit clusters in locations that qualitatively match the simulations, thereby supporting the assignment of these geometrical configurations.
The insets show ball-and-stick models of the corresponding molecular geometries.}
}
\end{figure}

\begin{figure}[h]
\includegraphics[width=0.8\columnwidth]{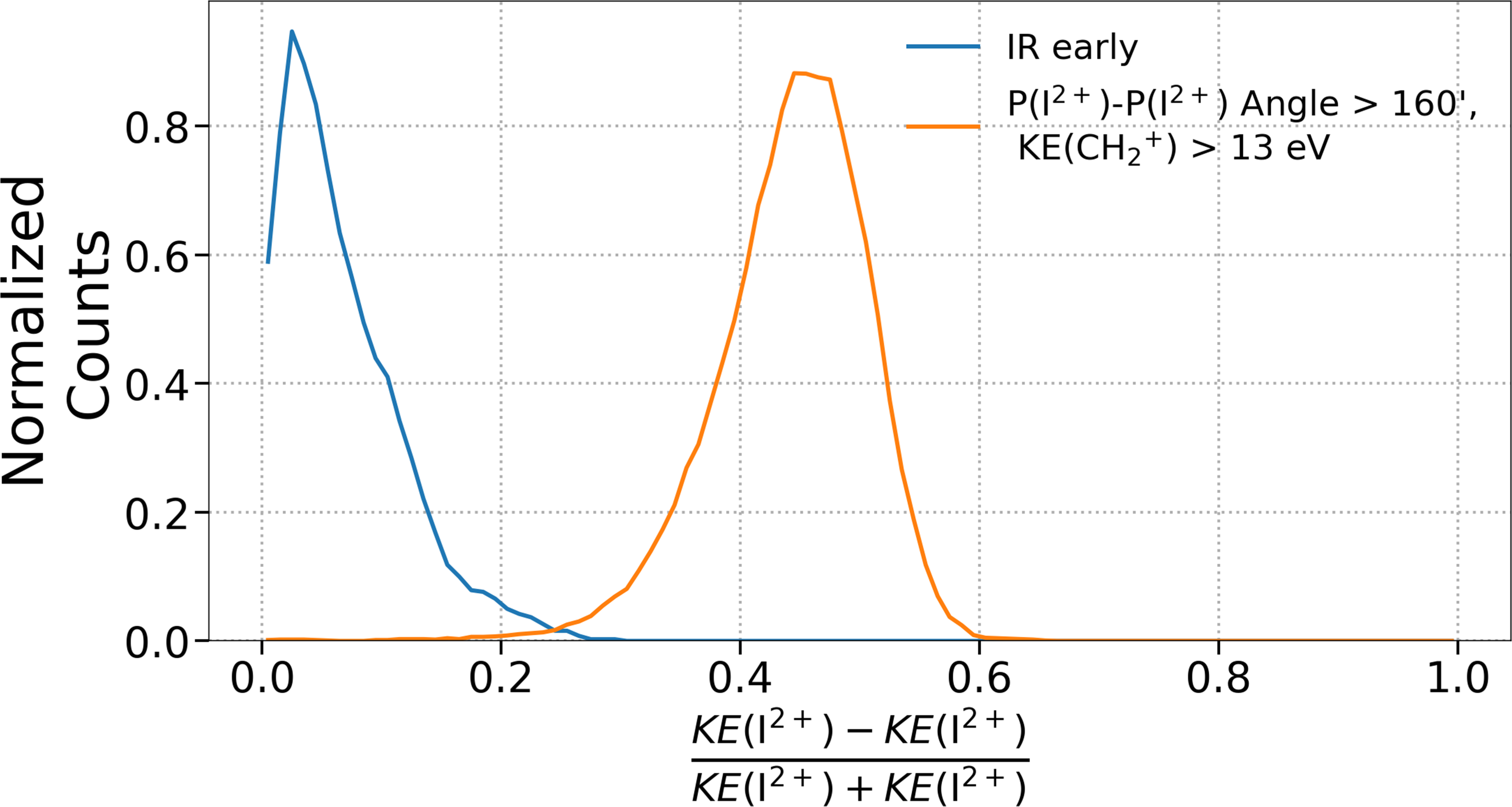}
\caption{\label{fig:SMEnergySharingRatio} \blue{Energy-sharing ratio between the two I$^{2+}$ fragments, defined as the normalized difference between the kinetic energies of the two I$^{2+}$ ions.  
For symmetric energy sharing, this ratio is expected to peak near zero, while asymmetric energy sharing produces broader distributions shifted away from zero.  
The blue curve corresponds to the IR-early events, where the distribution peaks near zero, indicating symmetric energy sharing consistent with the symmetric ground equilibrium geometry.  
The orange curve corresponds to the gated events shown in Fig.~\ref*{fig:KE2vsKE3SrtRng}(a), where the distribution is shifted and broadened around a peak near 0.5, revealing asymmetric energy sharing characteristic of these geometries. 
This comparison supports the interpretation of Fig.~\ref*{fig:KE2vsKE3SrtRng} that the iso-$\mathrm{CH_{2}I_{2}}$-like contribution is associated with asymmetric fragmentation dynamics.}
}
\end{figure}

\begin{figure}[h]
\includegraphics[width=\columnwidth]{Figures_SM/I-I_Distance_Estimation.png}
\caption{\label{fig:SM_II_distance}
\blue{(a) Calibration curve obtained from Coulomb explosion simulations with varied I--I separations, showing the correlation between the kinetic energy sum of the two I$^{2+}$ ions and the underlying I--I distance. 
The simulations were performed starting from the iso-geometry reported by Borin \textit{et al.}, and only the position of the outer I atom was varied along the I--I bond direction. 
(b) Estimated I--I distance distributions derived from the experimental data in Fig.~\ref*{fig:KE23SumSrtRng} using the calibration in panel (a). 
A Gaussian fit to the peak with high KE sum (corresponding to the green region in Fig.~\ref*{fig:KE23SumSrtRng}) yields a shortened I--I distance of $\sim$3.04~Å, while the Gaussian fit to the peak with low KE sum (red region in Fig.~\ref*{fig:KE23SumSrtRng}) yields an extended I--I distance of $\sim$4.56~Å.
The dashed line in black denotes the I--I distance in the neutral ground equilibrium CH$_{2}$I$_{2}$ geometry.
This analysis provides an approximate mapping between measured kinetic energies and transient I--I separation.}
}
\end{figure}

\clearpage 
\section{\label{sec:330nmPlots} Results for 330~\MakeLowercase{nm} excitation}

\blue{This section presents the comparative measurements performed with 330~nm UV excitation, complementing the 290~nm results discussed in the main text.  
The inclusion of this data set is motivated by earlier work from Borin et al.,\cite{Borin2016DirectStudy} who reported indications of transient iso-CH$_{2}$I$_{2}$-like configurations at 330~nm excitation in the gas phase.  
By analyzing our CEI observables under identical conditions, we are able to directly compare both excitation wavelengths.  
The results demonstrate that the signatures of transient isomer-like geometries appear at both 290~nm and 330~nm, with comparable relative yields and dynamics within the experimental uncertainties.  \\
}

\begin{figure}[h]
\includegraphics[width=0.8\columnwidth]{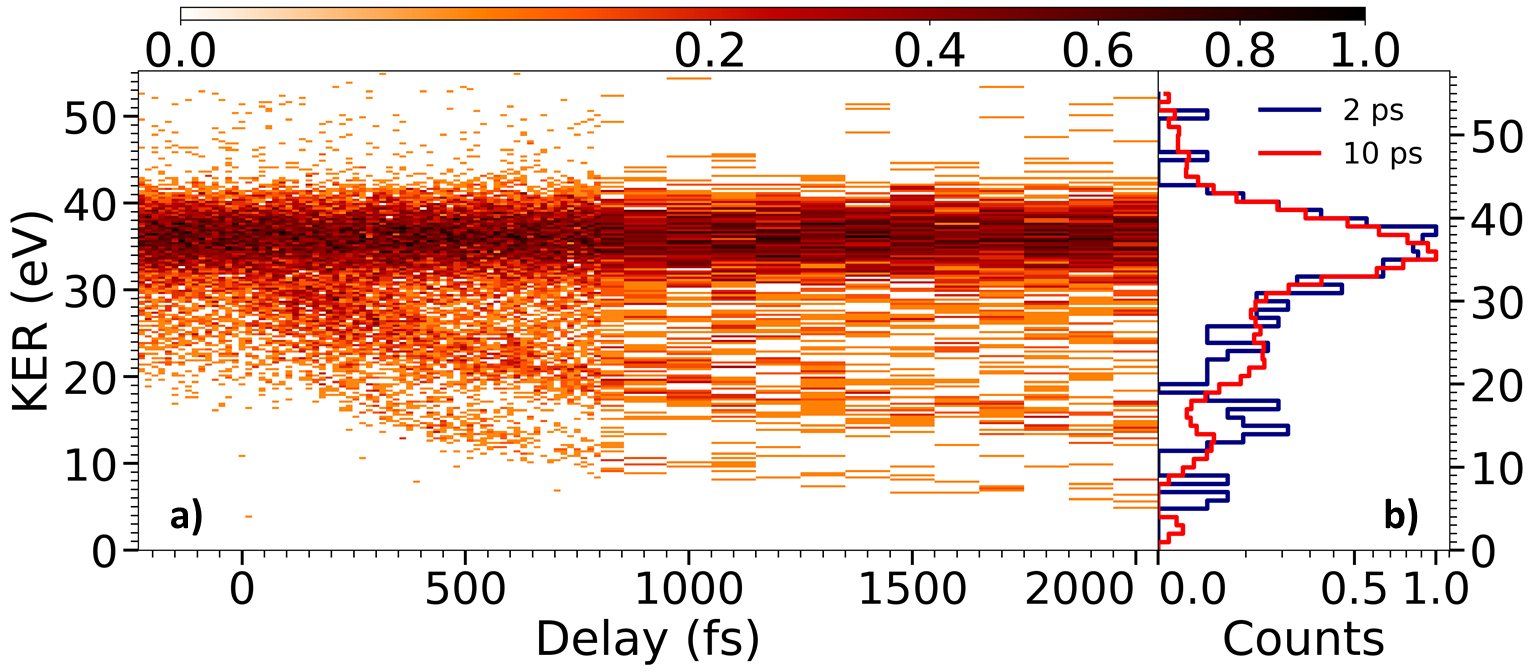}
\caption{\label{fig:SM330nmKERDelay} Coincidence ion yield of the (a) CH$_{2}^{+}$ + I$^{2+}$ + I$^{2+}$ channel as a function of pump-probe delay and KER. \textit{Right panel:} KER spectra at delays of 2~ps (blue) and 10~ps (red). The data shown here were recorded at a pump wavelength of 330~nm.}
\end{figure}

\begin{figure}
\includegraphics[width=0.8\columnwidth]{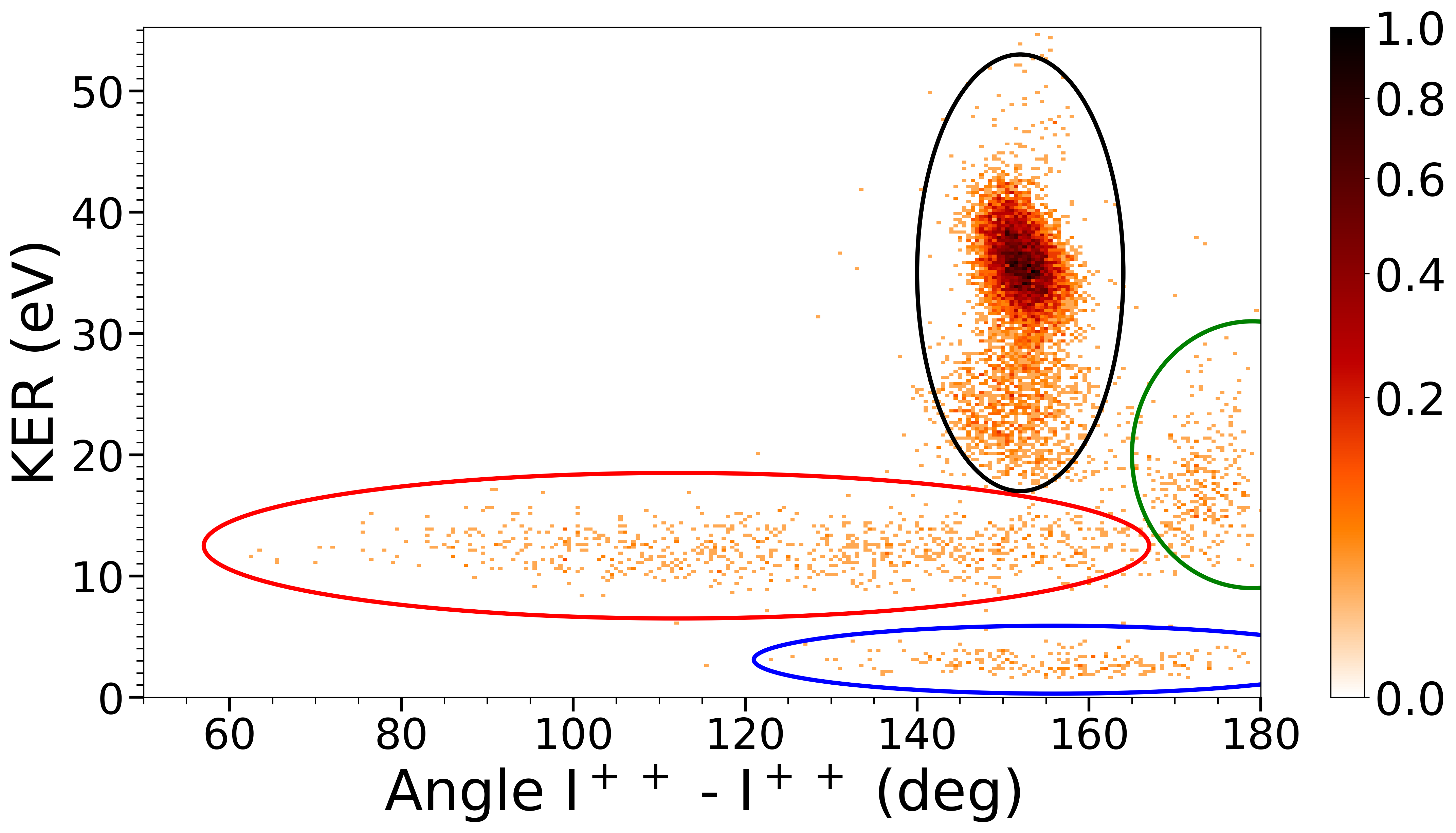}
\caption{\label{fig:SM330nmKERAng23} CH$_{2}^{+}$ + I$^{2+}$ + I$^{2+}$ ion yield as function of the KER and angle between the momentum vectors of the iodine ions, for the pump wavelength of 330~nm. The colored ovals mark the same contributions as described in Fig.~\ref*{fig:KERangAsym} of the main text.}
\end{figure}

\begin{figure}
\includegraphics[width=0.85\columnwidth]{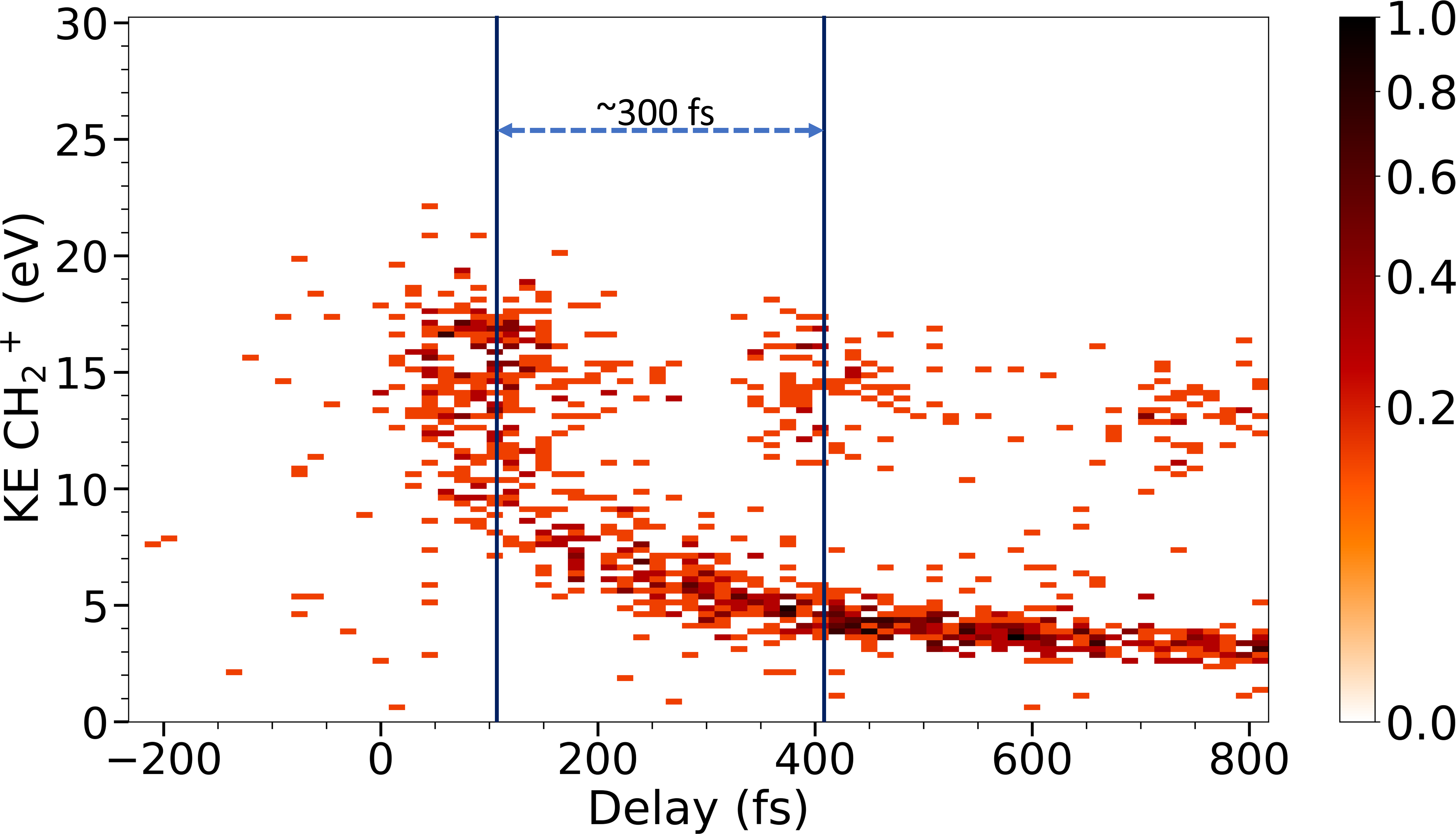}
\caption{\label{fig:SM330nmKE1delay_angGated} Yield of the CH$_{2}^{+}$ + I$^{2+}$ + I$^{2+}$ coincidence channel plotted as a function of pump--probe delay and the kinetic energy of the $\mathrm{CH_{2}^{+}}$ fragment, shown only for those events where the angle between the iodine ion momentum vectors is greater than 160$^\circ$.
\blue{The vertical lines indicate a time interval of $\sim$300~fs, corresponding to the rotational period of the neutral CH$_2$I radical formed after UV excitation and C--I bond cleavage. }
}
\end{figure}

\begin{figure}
\includegraphics[width=0.85\columnwidth]{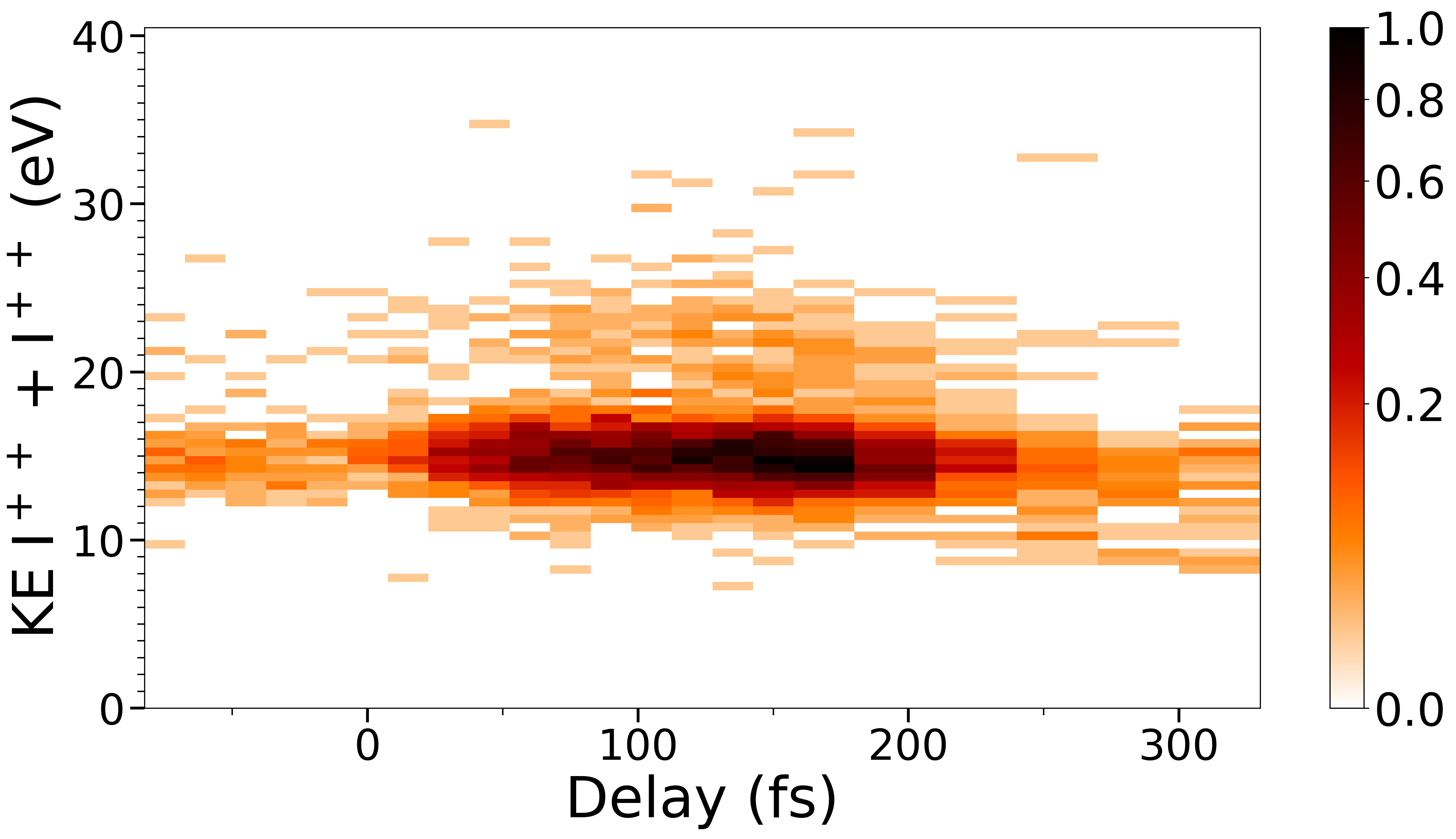}
\caption{\label{fig:SM330nmKEsum23_gated} Coincidence ion yield of the CH$_{2}^{+}$ + I$^{2+}$ + I$^{2+}$ channel as a function of the pump-probe delay and the kinetic energy sum of the two iodine ions, for events where the I--I momentum angle is greater than 160$^\circ$ and kinetic energy of $\mathrm{CH_{2}^{+}}$ is greater than 13~eV, similar to Fig.~\ref*{fig:KE23SumSrtRng} in the main text.}
\end{figure}

\begin{figure}
\includegraphics[width=0.9\columnwidth]{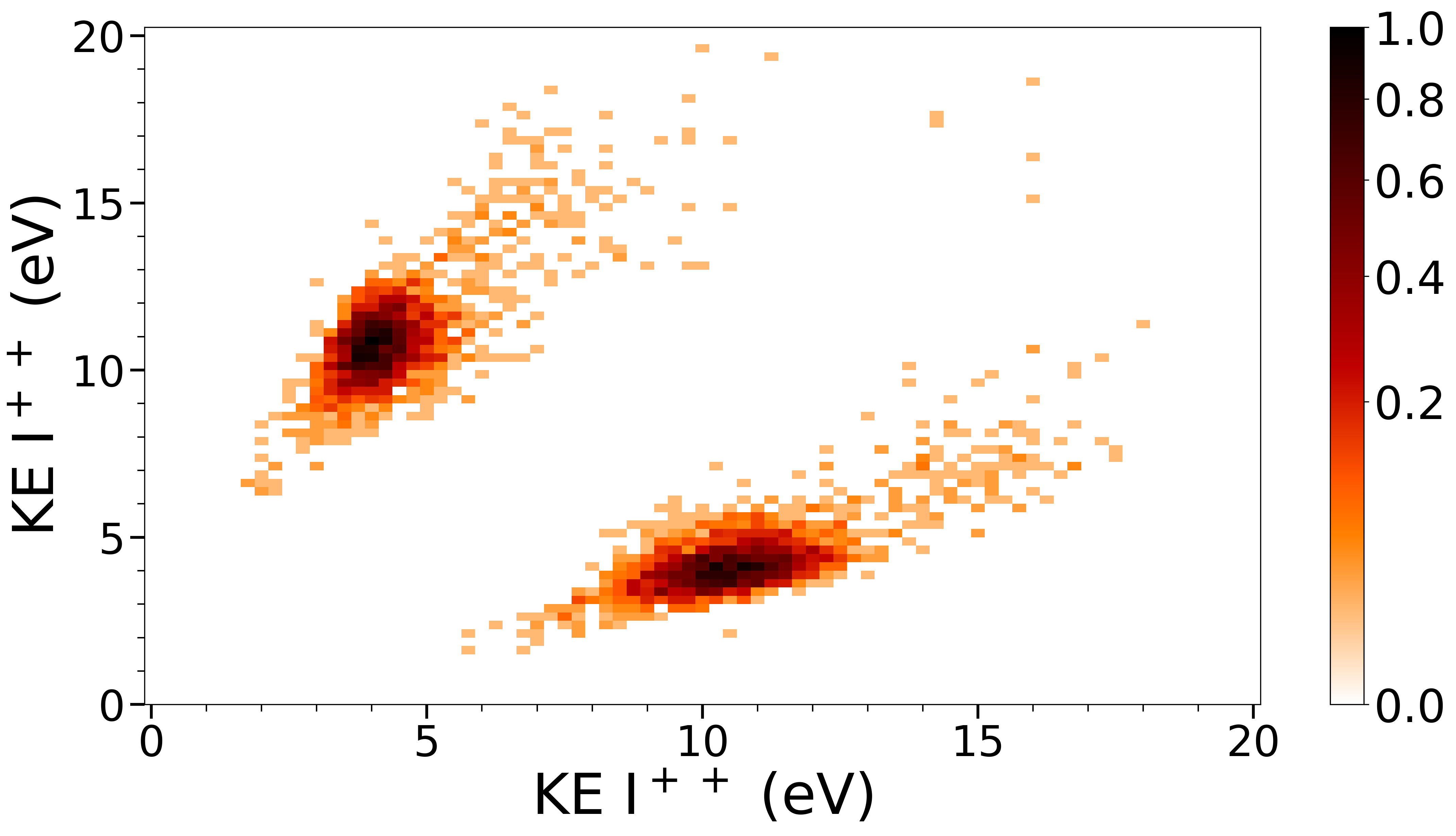}
\caption{\label{fig:SM330nmKE2vsKE3} Coincidence ion yield of the CH$_{2}^{+}$ + I$^{2+}$ + I$^{2+}$ channel as a function of the kinetic energy sharing between the two iodine ions. }
\end{figure}

\section{\label{sec:UVPowerDependence} UV Power Dependence}

\blue{This section presents the UV power dependence of the fragment yields extracted from different regions of the KER versus I--I angle distributions at long pump--probe delays.  
For both excitation wavelengths (290~nm and 330~nm), the log--log fits to the integrated yields reveal that the C--I dissociation channel scales with a slope of approximately one, consistent with a one-photon absorption process.  
By contrast, the molecular I$_{2}$ formation and the three-body C--I--I fragmentation channel exhibit slopes closer to two, indicating that these pathways proceed predominantly through a two-photon process.  
We also analyze the relative yields from the low-KE sum and high-KE sum regions, which both display slopes near one, confirming that these features arise from a one-photon excitation mechanism.  
Together, these power dependence measurements provide a clear distinction between single- and two-photon processes and further validate the channel assignments made in the main text.  }

\begin{figure}[h]
\includegraphics[width=\columnwidth]{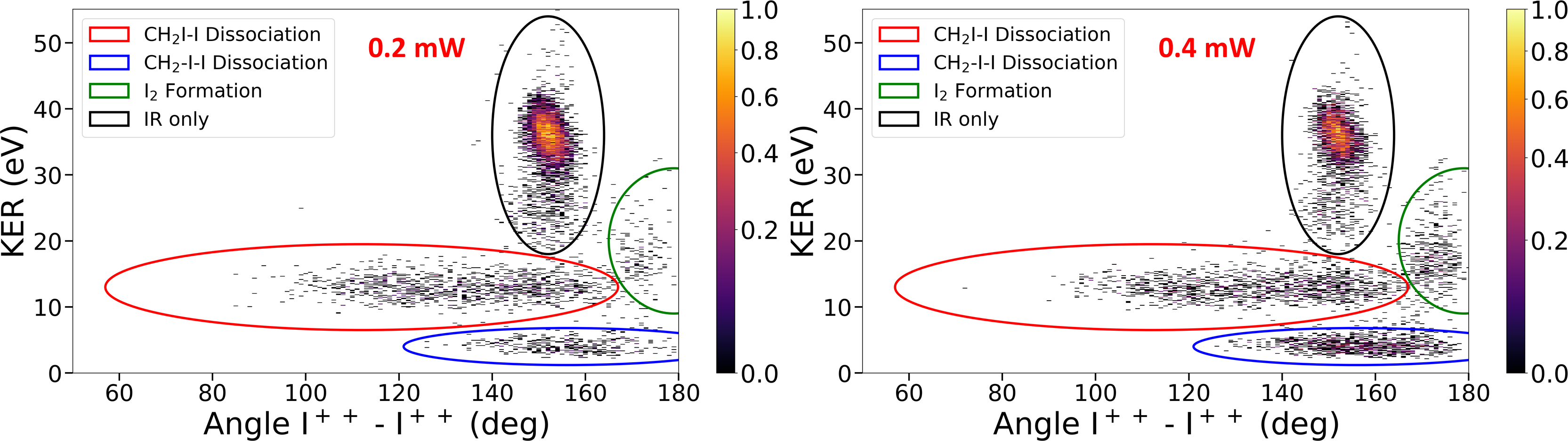}
\caption{\label{fig:SMUVpowerKERvsAng23} CH$_{2}^{+}$ + I$^{2+}$ + I$^{2+}$ ion yield as a function of KER and angle between the momenta of the iodine ions for two different powers of the 290~nm UV pulse. }
\end{figure}

\begin{figure}
\includegraphics[width=\columnwidth]{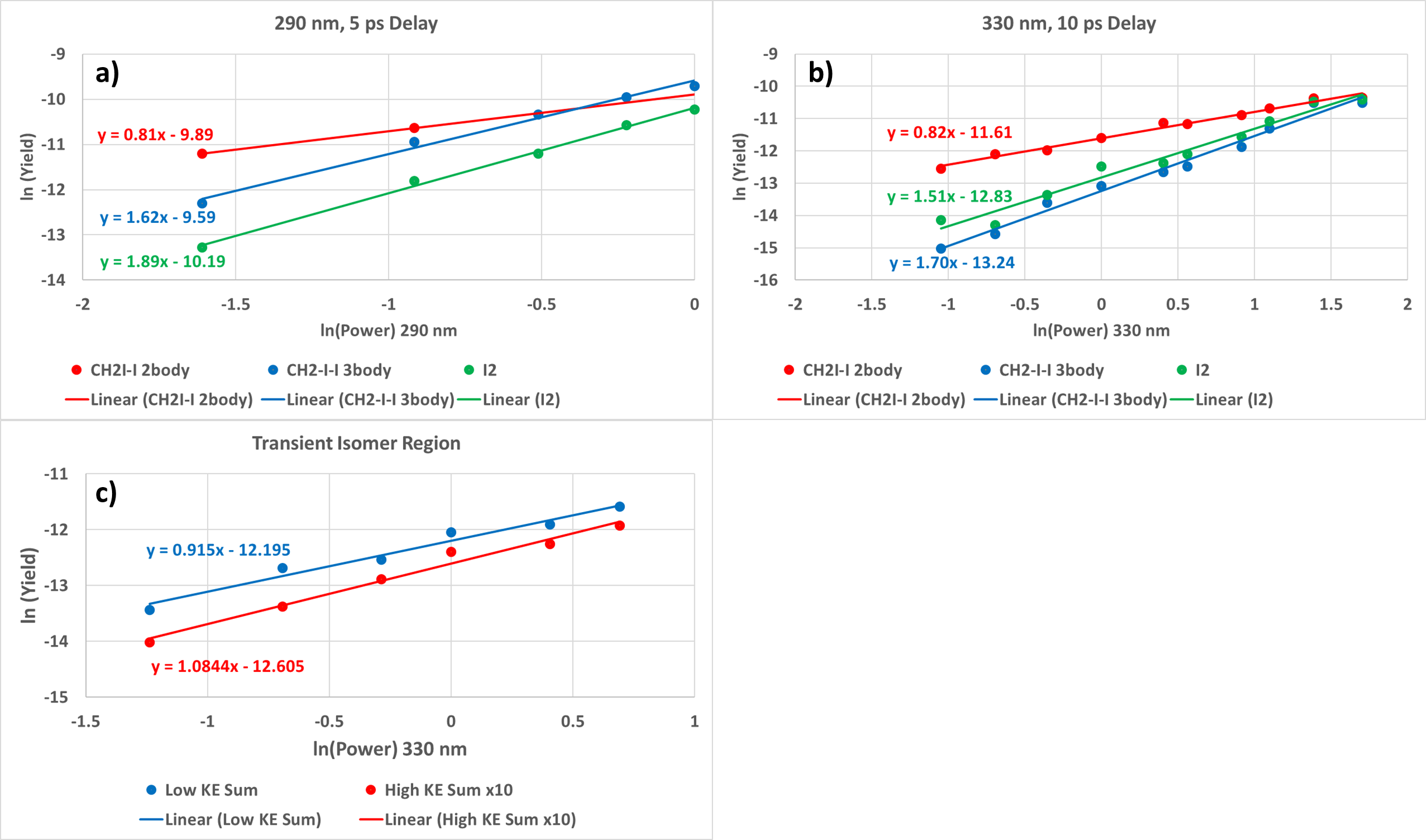}
\caption{\label{fig:SMUVpowerVSyield} Coincidence ion yields of different channels as a function of power of the 290~nm (a) and the 330~nm (b and c) UV pulses. The linear fits in the double logarithmic plot of yield vs average UV power give a rough estimate of the order of the UV-excitation process: a slope of approximately 1 corresponds to a single-photon process, a slope of approximately two to a double-photon process.}
\end{figure}

\section{\label{sec:Movie} Supplementary Movie}
Supplementary movie (Multimedia available online) shows the time evolution of the $\mathrm{CH_{2}^{+} + I^{2+} + I^{2+}}$ ion coincidence yield as a function of KER and angle between the momentum vectors of the two iodine dications (as plotted in Fig.~5 in the main text for a delay of 13.5 ps) for pump-probe delays between $-$200~fs (NIR before UV) and $+$2000~fs.

\begin{figure}
\includegraphics[width=0.95\columnwidth]{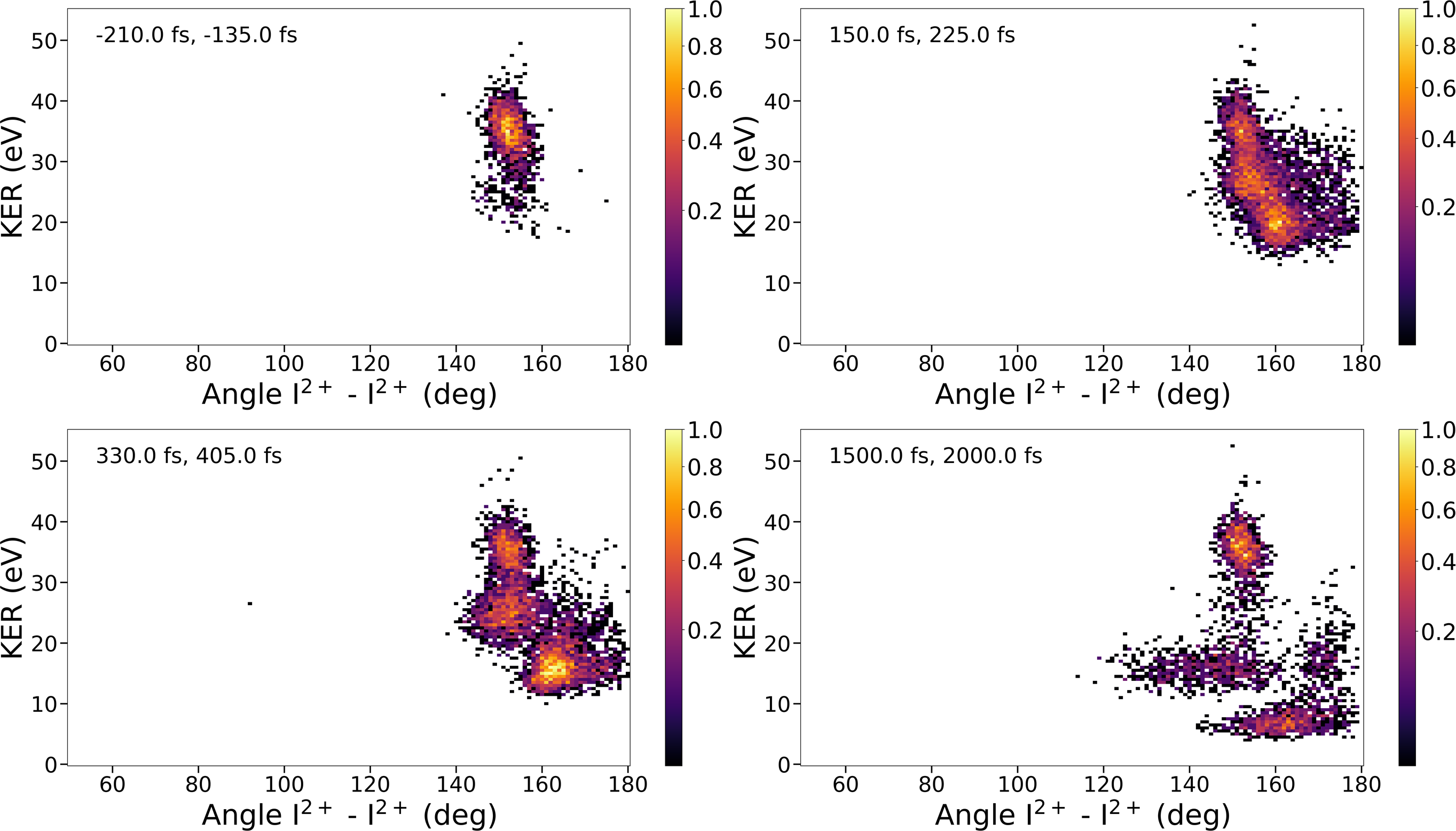}
\caption{\label{fig:SMKERMovStillImage} Representative snapshots of the movie (Multimedia available online) showing the time evolution.}
\end{figure}

\bibliography{CH2I2_UV_IR_CEI}